\documentclass[aps,prl,preprint,superscriptaddress]{revtex4-2}

\usepackage{amsmath,amssymb,amsfonts}
\usepackage{amsthm}
\usepackage{mathrsfs}

\usepackage{times}
\usepackage{color}

\usepackage{hyperref}
\hypersetup{hypertex=true,
	colorlinks=true,
	linkcolor=blue,
	anchorcolor=blue,
	citecolor=blue}

\usepackage{lineno}

\usepackage[defaultcolor=red]{changes}

\usepackage[normalem]{ulem} 

\begin{document}
	\title{Enhanced Anomalous Nernst Effect in the Ferromagnetic Kondo Lattice CeCo$_2$As$_2$}

\author{Shuyue Guan}
\thanks{These authors contribute equally.}
\affiliation{\label{aff1}International Center for Quantum Materials, School of Physics, Peking University, Beijing 100871, China}

\author{Weian Guo}
\thanks{These authors contribute equally.}
\affiliation{\label{aff2}School of Physics $\&$ Astronomy and Center for Advanced Quantum Studies, Beijing Normal University, Beijing 100875, China}

\author{Pengyu Zheng}
\affiliation{\label{aff2}School of Physics $\&$ Astronomy and Center for Advanced Quantum Studies, Beijing Normal University, Beijing 100875, China}

\author{Xinxuan Lin}
\affiliation{\label{aff1}International Center for Quantum Materials, School of Physics, Peking University, Beijing 100871, China}

\author{Yuqing Huang}
\affiliation{\label{aff1}International Center for Quantum Materials, School of Physics, Peking University, Beijing 100871, China}

\author{Jiawei~Li}
\affiliation{\label{aff1}International Center for Quantum Materials, School of Physics, Peking University, Beijing 100871, China}

\author{Xiao-Bin Qiang}
\affiliation{\label{aff3}State Key Laboratory of Quantum Functional Materials, Department of Physics, and Guangdong Basic Research Center of Excellence for Quantum Science, Southern University of Science and Technology, Shenzhen 518055, China}

\author{Longfei Li}
\affiliation{\label{aff1}International Center for Quantum Materials, School of Physics, Peking University, Beijing 100871, China}

\author{Weiwei~Xie}
\affiliation{\label{aff7}Department of Chemistry, Michigan State University, East Lansing, Michigan 48824, United States}

\author{Hai-Zhou Lu}
\affiliation{\label{aff3}State Key Laboratory of Quantum Functional Materials, Department of Physics, and Guangdong Basic Research Center of Excellence for Quantum Science, Southern University of Science and Technology, Shenzhen 518055, China}

\author{Zhiping Yin}
\email[]{yinzhiping@bnu.edu.cn}
\affiliation{\label{aff2}School of Physics $\&$ Astronomy and Center for Advanced Quantum Studies, Beijing Normal University, Beijing 100875, China}
\affiliation{\label{aff4}Key Laboratory of Multiscale Spin Physics (Ministry of Education), Beijing Normal University, Beijing 100875, China}

\author{Shuang Jia}
\email[]{gwljiashuang@pku.edu.cn}
\affiliation{\label{aff1}International Center for Quantum Materials, School of Physics, Peking University, Beijing 100871, China}
\affiliation{\label{aff5}Interdisciplinary Institute of Light-Element Quantum Materials and Research Center for Light-Element Advanced Materials, Peking University, Beijing 100871, China}
\affiliation{\label{aff6}Hefei National Laboratory, Hefei 230088, China}

\begin{abstract}
The anomalous Nernst effect (ANE), generating a voltage perpendicular to a temperature gradient due to magnetization, is closely linked to the Berry curvature (BC) near the Fermi energy in topological magnets. We report an enhanced spontaneous ANE in the ferromagnetic Kondo lattice CeCo$_2$As$_2$, which features Kondo-screened cerium-based $4f$ moments embedded in a ferromagnetic $d$-electron framework. The observed large anomalous Nernst coefficient, greater than the Seebeck coefficient, is attributed to the strong BC present in the $f$-orbital-dominated flat bands. The enhanced ANE in CeCo$_2$As$_2$ serves as a signature of the Fermi energy pinning within the topological flat band, highlighting the correlation-driven topology in the Kondo lattice.

\end{abstract}

\maketitle

\begin{figure}
	\includegraphics[width=1.0\linewidth]{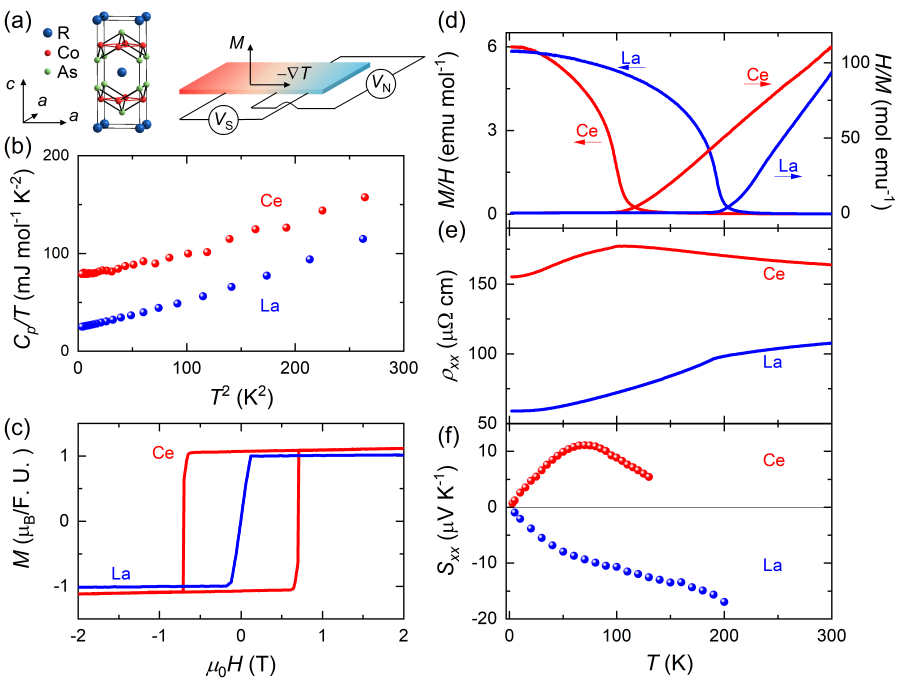} 
	\caption{Magnetization, specific heat, and transport properties of CeCo$_2$As$_2$ and LaCo$_2$As$_2$. (a) Unit cell of CeCo$_2$As$_2$ ($a=4.03$~\AA, $c=10.20$~\AA) and schematic illustrations of the Seebeck effect and the ANE. The Seebeck coefficient is defined as $S_{xx}=E_{x}/\nabla_{x}T$ while the anomalous Nernst coefficient is defined as $S^{\mathrm{A}}_{xy}=-E_{y}^{\mathrm{A}}/\nabla_{x}T$. The spontaneous magnetization $M$ is along the $c$ axis, and the heat current is applied along the $a$ axis. The anomalous Nernst voltage $V^\mathrm{A}_{\mathrm{N}}$ appears perpendicular to both the magnetization and the temperature gradient. In transport measurements, the $x$, $y$, and $z$ axes are defined along the crystallographic $a$, $a$, and $c$ axes, respectively. (b) The low-temperature specific heat divided by the temperature [$C_p(T)/T$] of CeCo$_2$As$_2$ and LaCo$_2$As$_2$, indicating that $C_p(T)$ can be described as $C_p(T) = \gamma T + \beta T^3$, where $\gamma T$ and $\beta T^3$ are the electronic and the lattice contribution, respectively. (c) The magnetic field dependence of the magnetization at 2~K. (d)-(f) The temperature dependence of the $M/H$ and $H/M$ curves in an external magnetic field of $0.1$ T applied in the crystallographic $c$ axis, the longitudinal electric resistivity $\rho_{xx}(T)$, and the Seebeck coefficient $S_{xx}(T)$. } 
	\label{Figure1} 
\end{figure}

The anomalous Nernst effect (ANE) acts as a thermoelectric counterpart to the anomalous Hall effect (AHE). It holds promise for thermoelectric devices by facilitating perpendicular current flow and improving design and performance \cite{deviceNakatsuji2019, Fu2020TEP}.
Unlike the conventional Nernst effect, which necessitates an external magnetic field, the ANE in a ferromagnetic (FM) material emerges intrinsically from the magnetization, allowing for heat-to-spin/charge conversion devoid of an external magnetic field [Fig.~\ref{Figure1}(a)].
However, the anomalous Nernst coefficient ($S^{\mathrm{A}}_{xy}=-E_{y}^{\mathrm{A}}/\nabla_{x}T$) in conventional magnets typically measures between $0.1-1$ $\mu$V K$^{-1}$, which is typically smaller than the Seebeck coefficient by several orders of magnitude, limiting their applicability \cite{Ramos2014Fe3O4,Hasegawa2015Fe,Chuang2017FeCoNi}. 

It is well-established that large intrinsic AHE and ANE are distinguishing traits of topological magnets, heavily reliant on the Berry curvature (BC) distribution around the Fermi level ($E_{\mathrm{F}}$) \cite{BerryXiao2006, Mn3SnIkhlas2017,Co2MnGaSakai2018,Co3Sn2S2Guin2018,Fe3GeTe2Xu2019,Fe3GaSakai2020,Mn3GeXu2020,Fe3Sn2Zhang2021,UCoRuAlAsaba2021,YbMnBi2Pan2022,Fe3SnChen2022,CeCrGe3Li2024}.
High-throughput first-principles calculations have predicted magnetic materials with notable topological structures, including nodal lines and Weyl points, where   BC can yield greater $S^{\mathrm{A}}_{xy}$ \cite{Noky2018Ti2MnAl,Noky2019Fe2MnX,Noky2020Heusler,Fe3GaSakai2020}.
Recent experimental work demonstrates that the large ANE is highly sensitive to $E_{\mathrm{F}}$ in the topological magnets \cite{Li2023Fe3Sn2,Liu2023Co3Sn2S2FeNi}. 

In this Letter, we underscore that the $f$-electron Kondo effect can significantly boost the ANE by generating a narrow topological band through $f$-$d$ hybridization.
We conduct a comparative analysis of the magnetization, electric, and thermoelectric properties of CeCo$_2$As$_2$ and its iso-structural, non-$4f$ analog LaCo$_2$As$_2$, both belonging to the $R$Co$_2$As$_2$ family ($R$ = light rare earth), crystallizing in the ThCr$_2$Si$_2$-type structure (space group $I4/mmm$).
Their magnetic characteristics are shaped by the interplay of itinerant Co $d$ electrons and localized $R$ $4f$ electrons \cite{CCAThompson2014,CCATan2018,Huang2023anomalous}. 
Both CeCo$_2$As$_2$ and LaCo$_2$As$_2$ exhibit an FM ground state with the $c$ axis as the easy axis and the Curie temperature ($T_{\mathrm{C}}$) above 100~K~\cite{CCAThompson2014,CCATan2018}. While LaCo$_2$As$_2$ acts as a $d$-electron magnet, CeCo$_2$As$_2$ emerges as an FM Kondo lattice in which the Kondo effect is pivotal for Ce's $4f$ electron behavior, forming topological bands with pronounced $f$-orbital character \cite{CCAcheng2023}.   

Our research unveils a unique topological thermoelectric signal in the single crystals of CeCo$_2$As$_2$, an impressive value of $S^{\mathrm{A}}_{xy}$ of $7.4$ $\mu$V K$^{-1}$ at 40~K.
Furthermore, the values of $S^{\mathrm{A}}_{xy}$ surpass those of Seebeck coefficient $S_{xx}$ across a broad temperature range, leading to an anomalous Nernst angle ($\tan\theta_{\mathrm{AN}} = \left| S^{\mathrm{A}}_{xy}/S_{xx} \right|$) as large as 144\%.
Our calculations propose that the large AHE and ANE arise from the pronounced BC in the $f$-electron characteristic bands, with multiple hybridization gaps and Weyl nodes existing within several meV of $E_{\mathrm{F}}$.
As the Kondo hybridization effectively pins $E_{\mathrm{F}}$ within these topological flat bands, the ANE is further amplified.
These findings offer a unique perspective on the potential of correlated topological magnets in thermoelectric applications, providing insight into the critical interplay between the Kondo effect and electronic topology.

Single crystals of CeCo$_2$As$_2$ and LaCo$_2$As$_2$ were synthesized via a self-flux technique \cite{Huang2023anomalous}, and their thermodynamic and transport properties are presented in  Fig.   \ref{Figure1}.
The electronic specific heat coefficient ($\gamma$) of CeCo$_2$As$_2$ is measured to be $78\ \mathrm{mJ\ mol^{-1}\ K^{-2}}$, which is approximately three times greater than that of LaCo$_2$As$_2$, indicating a significant enhancement in the effective mass of charge carriers in CeCo$_2$As$_2$.
The temperature dependence of longitudinal resistivity  [$\rho_{xx}$($T$)] of CeCo$_2$As$_2$ demonstrates an initial increase with decreasing temperature, culminating in a broad maximum. 
This behavior signifies a transition from a regime dominated by incoherent scattering of the carriers on $f$ local moments at elevated temperatures to a state characterized by coherent scattering at lower temperatures for a Kondo lattice \cite{HFbook2007,KondoscaYang2008,Jang2020evoKL}. The broad maximum in $\rho_{xx}$($T$) signifies this transition with a characteristic coherence temperature ($T_{\mathrm{coh}}\sim 100 $~K) for CeCo$_2$As$_2$.
In addition, CeCo$_2$As$_2$ presents a positive $S_{xx}$ with a peak at 70 K.
Large and positive $S_{xx}$ is commonly observed in cerium-based heavy fermions (HFs) due to the intricate interplay between the Kondo effect and the crystalline electric field \cite{SgammaBehnia2004}.
Conversely, LaCo$_2$As$_2$ exhibits normal metallic behavior in its resistivity and a negative $S_{xx}$.

It is noteworthy that CeCo$_2$As$_2$ and LaCo$_2$As$_2$ exhibit similarities in their temperature and field-dependent magnetization.
The temperature dependence of their susceptibilities aligns with Curie-Weiss behavior at elevated temperatures, with fitting yielding an effective magnetic moment $\mu_{\mathrm{eff}}=3.6~ \mu_{\mathrm{B}}$/F. U. for CeCo$_2$As$_2$.
This value is close to the combined $\mu_{\mathrm{eff}}$ for LaCo$_2$As$_2$ ($2.9~ \mu_{\mathrm{B}}$/F. U. ) and the Hund's rule ground state moment for the Ce$^{3+}$ ion ($2.54~ \mu_{\mathrm{B}}$/ Ce$^{3+}$), indicating a decoupling of the $4f$ local moment from itinerant electrons at high temperatures.
The compounds display FM ordering at $T_{\mathrm{C}}$ of 190~K and 99~K, with a marked increase in magnetization observed under a weak external magnetic field below $T_{\mathrm{C}}$.
Furthermore, a change in the slope of the temperature-dependent resistivity is noted at $T_{\mathrm{C}}$, attributed to the suppression of spin disorder scattering (See more details in Supplementary Materials~\cite{SM} for the determination of $T_{\mathrm{C}}$ and $T_{\mathrm{coh}}$, available in the published version).
The lower $T_{\mathrm{C}}$ of CeCo$_2$As$_2$ relative to $T_{\mathrm{coh}}$ suggests that its FM state involves collective Kondo hybridization.
The nearly identical saturated magnetization of $1.0~\mu_{\mathrm{B}}$/ F. U. for both compounds at 2~K, as depicted in Fig. \ref{Figure1}(c), supports the conclusion that the $4f$ local moment of the cerium atom at low temperatures is effectively quenched.

\begin{figure}
	\includegraphics[width=1\linewidth]{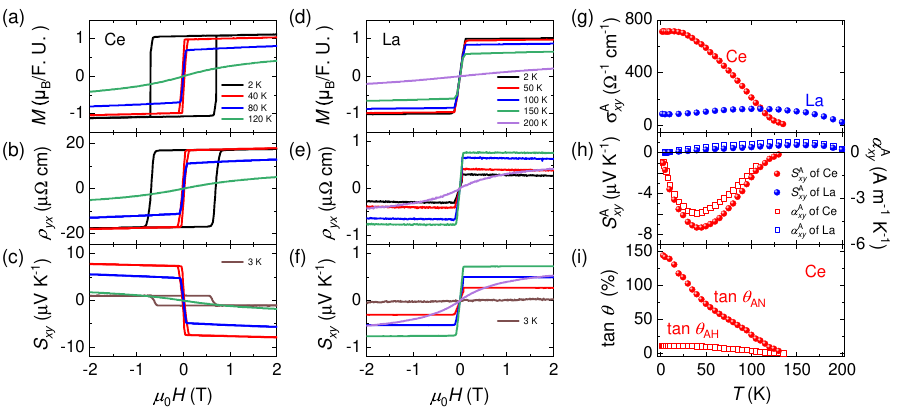} 
	\caption{The magnetic field dependence of the magnetization $M$, the Hall resistivity $\rho_{yx}$ and the Nernst coefficients $S_{xy}$ of (a)-(c) CeCo$_2$As$_2$, and (d)-(f) LaCo$_2$As$_2$ at several representative temperatures. The temperature dependence of (g) $\sigma_{xy}^{\mathrm{A}}$ and (h) $S_{xy}^{\mathrm{A}}$ and $\alpha_{xy}^{\mathrm{A}}$ of CeCo$_2$As$_2$ and LaCo$_2$As$_2$. (i) The temperature dependence of $\tan\theta_{\mathrm{AH}}$ and $\tan\theta_{\mathrm{AN}}$ of CeCo$_2$As$_2$.} 
	\label{Figure2} 
\end{figure}

The Hall resistivity $\rho_{yx}$($H$) and the Nernst signal $S_{xy}$($H$), as illustrated in Figs.~\ref{Figure2}(b) and (e), and Figs.~\ref{Figure2}(c) and (f), respectively, exhibit profiles that are closely comparable to the magnetization curves, as shown in Figs.~\ref{Figure2}(a) and (d).
Notably, CeCo$_2$As$_2$ displays significantly higher absolute values for both $\rho_{yx}$ and $S_{xy}$ than LaCo$_2$As$_2$.
In FM conductors, the Hall resistivity can be decomposed into ordinary and anomalous contributions, expressed as $\rho_{yx} = \rho^{\mathrm{O}}_{yx}+\rho^{\mathrm{A}}_{yx}$, where $\rho^{\mathrm{O}}_{yx}$ and $\rho^{\mathrm{A}}_{yx}$ correspond to ordinary and anomalous Hall resistivities, respectively, which are linearly dependent on the external magnetic field and magnetization \cite{RevAHE2010}.
Similarly, the Nernst signal can also be categorized into ordinary and anomalous contributions: $S_{xy} = S^{\mathrm{O}}_{xy}+S^{\mathrm{A}}_{xy}$.
Extrapolation of the high-field regions of $\rho_{yx}$ and $S_{xy}$ to the zero-field limit indicates that  $\rho^{\mathrm{A}}_{yx}$ and $S^{\mathrm{A}}_{xy}$ of CeCo$_2$As$_2$ are an order of magnitude greater than those for LaCo$_2$As$_2$.

The temperature dependence of crucial parameters associated with anomalous effects, including the anomalous Hall conductivity $\sigma^{\mathrm{A}}_{xy} \simeq \rho^{\mathrm{A}}_{yx}/\rho_{xx}^2$, the anomalous Nernst coefficient $S^{\mathrm{A}}_{xy}$, and the anomalous Nernst conductivity $\alpha^{\mathrm{A}}_{xy} = \sigma^{\mathrm{A}}_{xy} S_{yy} + \sigma_{xx} S^{\mathrm{A}}_{xy}$ for both CeCo$_2$As$_2$ and LaCo$_2$As$_2$, is summarized in  Figs.~\ref{Figure2}(g) and (h).
As the temperature decreases, $\sigma^{\mathrm{A}}_{xy}$ of CeCo$_2$As$_2$ increases monotonically and reaches 710~$\Omega^{-1}$ cm$^{-1}$ below 20~K, significantly exceeding the maximum value observed in LaCo$_2$As$_2$, which is 130~$\Omega^{-1}$\,cm$^{-1}$ at 110~K.
Consequently, anomalous Hall angle $\tan\theta_{\mathrm{AH}} = \left|\sigma^{\mathrm{A}}_{xy}/\sigma_{xx} \right|$ for CeCo$_2$As$_2$ attains its maximum of 11\% at 2~K, as shown in Fig.~\ref{Figure2}(i).
The values of $S^{\mathrm{A}}_{xy}$ for CeCo$_2$As$_2$ remain negative, with the absolute value increasing with temperature, peaking at 7.4 $\mu$V K$^{-1}$ around 40~K. In contrast, the values of $S^{\mathrm{A}}_{xy}$ for LaCo$_2$As$_2$ are positive and gradually rise with temperature, reaching 0.7 $\mu$V K$^{-1}$ at approximately 150~K.
The observed peak value of $S^{\mathrm{A}}_{xy}$ for CeCo$_2$As$_2$ is comparable to some of the largest values reported for other topological magnets, such as 6 $\mu$V K$^{-1}$ for Co$_2$MnGa \cite{Co2MnGaSakai2018,Guin2019Co2MnGa, Xu2020Mott}, 3-5 $\mu$V K$^{-1}$ for Co$_3$Sn$_2$S$_2$ \cite{Co3Sn2S2Guin2018,Ding2019mobility,Yang2020Co3Sn2S2}, and 6-10 $\mu$V K$^{-1}$ for YbMnBi$_2$ \cite{YbMnBi2Pan2022, Guo2023Onsager}.
The temperature dependence of $\alpha^{\mathrm{A}}_{xy}$ of CeCo$_2$As$_2$ follows the profile of that of $S^{\mathrm{A}}_{xy}$, with a peak value of $4.0$ A m$^{-1}$ K$^{-1}$, ranking among the top values reported in topological magnets \cite{Co2MnGaSakai2018,Co3Sn2S2Guin2018,Fe3GaSakai2020,UCoRuAlAsaba2021,YbMnBi2Pan2022,Fe3SnChen2022,CeCrGe3Li2024}. In contrast, LaCo$_2$As$_2$ exhibits positive $\alpha^{\mathrm{A}}_{xy}$ with a small peak value of $0.7$ A m$^{-1}$ K$^{-1}$.
Notably, the absolute values of $S^{\mathrm{A}}_{xy}$ and $S_{xx}$ for CeCo$_2$As$_2$ are comparable below its $T_{\mathrm{C}}$, which emphasizes a significant $\tan\theta_{\mathrm{AN}}$ of 87\% at 40 K, coinciding with the maximum of $S^{\mathrm{A}}_{xy}$.
As shown in Fig.~\ref{Figure2}(i), the values of $\tan\theta_{\mathrm{AN}}$ exceed 100\% below 30 K, ultimately reaching 144\% at the lowest measured temperature of 3~K.
In comparison, the $\tan\theta_{\mathrm{AN}}$ values for conventional bulk magnets such as Fe, Co, and Ni remain below 1\% \cite{Chuang2017FeCoNi}.

\begin{figure}
	\includegraphics[width=1\linewidth]{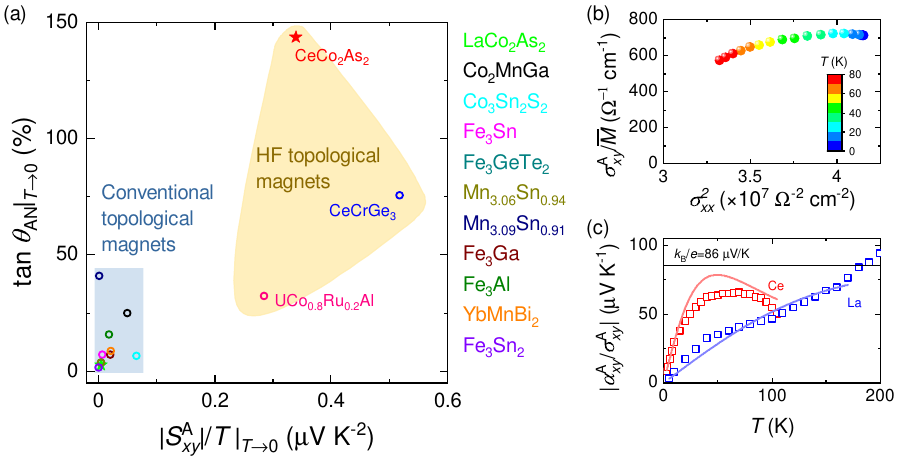} 
	\caption{(a) Summary of the data of $\tan\theta_{\mathrm{AN}}$ and $\left|S^{\mathrm{A}}_{xy}/T\right|$ at the lowest temperature for various topological magnets. (b) Scaling of $\sigma^{\mathrm{A}}_{xy}$ for CeCo$_2$As$_2$ by the normalized magnetization $\overline{M}=M(T)/{M_{\mathrm{sat}}}(T=2\mathrm{\ K})$ with $\sigma^2_{xx}$. (c) The $\left|\alpha^{\mathrm{A}}_{xy}/\sigma^{\mathrm{A}}_{xy}\right|$ ratio as a function of temperature for CeCo$_2$As$_2$ and LaCo$_2$As$_2$. The solid lines are the fitting curves.} 
	\label{Figure3} 
\end{figure}

To illustrate the substantial ANE for CeCo$_2$As$_2$, we summarize $\tan\theta_{\mathrm{AN}}$ and $\left|S^{\mathrm{A}}_{xy}/T\right|$ at the lowest temperatures for various topological magnets \cite{Mn3SnIkhlas2017,Co2MnGaSakai2018,Co3Sn2S2Guin2018,Fe3GeTe2Xu2019,Fe3GaSakai2020,Mn3GeXu2020,Fe3Sn2Zhang2021,UCoRuAlAsaba2021,YbMnBi2Pan2022,Fe3SnChen2022,CeCrGe3Li2024}, as depicted in Fig.~\ref{Figure3}(a).
In a multiband system, $S_{xx}$ can vanish due to electron–hole compensation at finite temperatures, in which case the $\tan\theta_{\mathrm{AN}}$ may take arbitrarily large values~\cite{Mn3SnIkhlas2017, Fe3Sn2Zhang2021, UCoRuAlAsaba2021}.
However, we note that the values of  $\tan\theta_{\mathrm{AN}}$ for most of the magnets are less than 50\% at low temperatures in general.
Correspondingly, the values of  $\left|S^{\mathrm{A}}_{xy}/T\right|$ are less than $0.1~\mu$V K$^{-2}$.
Three $f$-electron compounds, UCo$_{0.8}$Ru$_{0.2}$Al \cite{UCoRuAlAsaba2021}, which shows great value of $S^{\mathrm{A}}_{xy}$ being 23 $\mu$V K$^{-1}$ at 40~K, CeCrGe$_3$ \cite{CeCrGe3Li2024}, and CeCo$_2$As$_2$, exhibit substantial values of $S^{\mathrm{A}}_{xy}/T$.

It is noteworthy that these three $f$-electron compounds can be classified as HFs, which exhibit enhanced values of $\gamma$ at low temperatures.
On the other hand, transport measurements on CeCo$_2$As$_2$ provide clear evidence of its single-band nature and reveal a large $S_{xx}$, making the exceptionally large $\tan\theta_{\mathrm{AN}}$ remarkable (See details in the Supplementary Materials~\cite{SM}, available in the published version).
Because the large $S_{xx}$ at low temperatures reflects the significant density of states (DOS) in HF compounds~\cite{SgammaBehnia2004}, the large value of $\tan\theta_{\mathrm{AN}}$ should be ascribed to a large $S^{\mathrm{A}}_{xy}/T$ rather than a small $S_{xx}$, highlighting the significant ANE in HF magnets.

We scale $\sigma^{\mathrm{A}}_{xy}$ for CeCo$_2$As$_2$ divided by its normalized magnetization $\overline{M}=M(T)/{M_{\mathrm{sat}}}(T=2\mathrm{\ K})$ against $\sigma^2_{xx}$ in Fig. \ref{Figure3}(b), taking into account the variation in magnetization from 2 K to $T_{\mathrm{C}}$.
The AHE may arise from either extrinsic skew scattering contribution, leading to a quadratic relationship between the $\sigma^{\mathrm{A}}_{xy}$ and $\sigma_{xx}$ \cite{tian2009proper}, or from an intrinsic BC contribution, wherein $\sigma^{\mathrm{A}}_{xy}$ remains independent of $\sigma_{xx}$ \cite{Onoda2008quantum, RevAHE2010}.
The values of $\sigma^{\mathrm{A}}_{xy}/\overline{M}$ in CeCo$_2$As$_2$ consistently hover around  700 $\Omega^{-1}$ cm$^{-1}$, indicating a prevailing intrinsic contribution.
It was reported that the skew scattering AHE in HFs, due to the disrupted Kondo resonance of the local moment scatters conduction electrons in a magnetic field \cite{Nair2012HEinHF, Yang2016twofluid}, is characterized by a small value of $\tan\theta_{\mathrm{AH}}$, being approximately 1\% \cite{RevAHE2010}.
The exceptional ANE and AHE observed in CeCo$_2$As$_2$  necessitate a reevaluation of the role of the Kondo effect in its topological electronic structure.

\begin{figure}
	\includegraphics[width=1\linewidth]{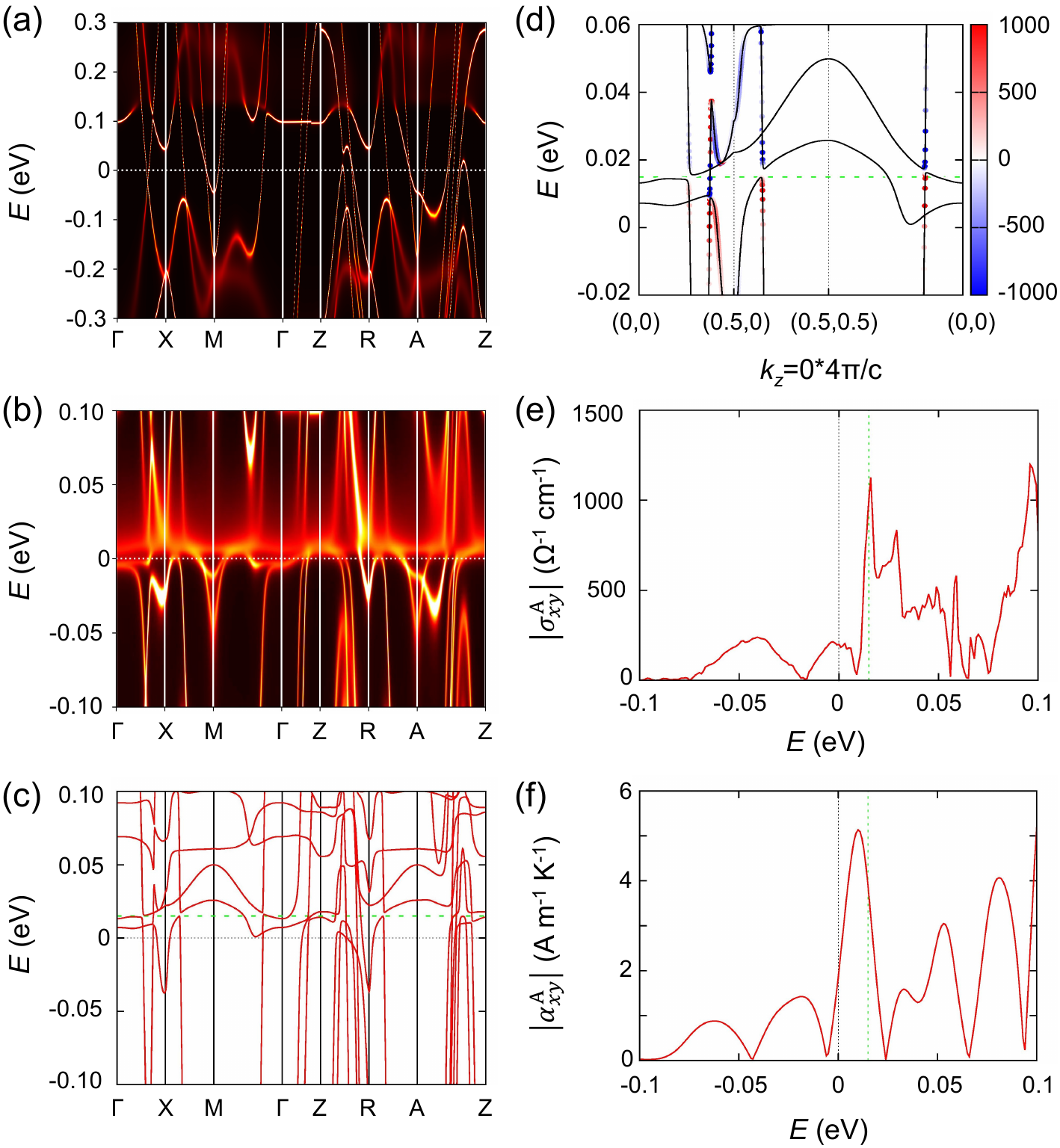} 
	\caption{The DFT+DMFT band structures of (a) LaCo$_2$As$_2$ and (b) CeCo$_2$As$_2$, respectively; (c) the band structure of CeCo$_2$As$_2$ obtained from the adjusted DFT TB model; and, based on (c), (d) the Berry-curvature–resolved band structure with the $z$-component $\Omega_z$ (\AA$^2$) in the $k_z=0*4\pi/c$ plane as a representative result, (e) the calculated absolute $\sigma^{\mathrm{A}}_{xy}$, and (f) the calculated absolute $\alpha^{\mathrm{A}}_{xy}$ at $T=50$~K.  In (d)-(f), the green dashed lines represent $E$=$E_{\mathrm{F}}$+15 meV.}
	\label{Figure4} 
\end{figure}

To illuminate the role of $4f$ electrons in the band structure, we conducted a comparison of LaCo$_2$As$_2$ and CeCo$_2$As$_2$ utilizing density functional theory coupled with dynamical mean-field theory (DFT+DMFT), as illustrated in Figs.~\ref{Figure4}(a) and (b), respectively.
Significant Kondo flat bands are observed just above $E_{\mathrm{F}}$ in CeCo$_2$As$_2$, which are absent in LaCo$_2$As$_2$.
The presence of these Kondo flat bands elucidates the HF properties and the significant value of $S_{xx}/T\vert_{T\rightarrow 0}$, as the latter is proportional to the DOS at $E_{\mathrm{F}}$.
Further analysis reveals numerous $f$-$d$ hybridization gaps and Weyl points distributed within a narrow energy range near $E_{\mathrm{F}}$  (See Fig. S13 in the Supplementary Material~\cite{SM}, available in the published version).

Because direct determination of the BC is not feasible in DFT+DMFT calculations, we employed the adjusted DFT tight-binding (TB) model to capture the characteristics of Kondo flat bands and their hybridization with Co-$3d$ bands, as shown in Fig.~\ref{Figure4}(c). 
A representative BC–resolved band structure with the $z$-component $\Omega_z$ (\AA$^2$) in the $k_z=0*4\pi/c$ plane, along with those for other planes, is shown in Fig.~\ref{Figure4}(d) and the Supplementary Materials~\cite{SM}, available in the published version.
Our findings demonstrate that the large $\sigma^{\mathrm{A}}_{xy}$ and $\alpha^{\mathrm{A}}_{xy}$ can be attributed to the significant average BC arising from the $f$-$d$ hybridization gaps and associated Weyl points. These features not only increase $\sigma^{\mathrm{A}}_{xy}$ \cite{Sun2018away}, as the three-dimensional Kondo flat band lies near $E_{\mathrm{F}}$, amplifying the BC contribution from occupied states across all $k_z$ planes (Fig. S15 in the Supplementary Material~\cite{SM}, available in the published version), but also enhance $\alpha^{\mathrm{A}}_{xy}$ by concentrating a large average BC within a narrow energy window near the $E_{\mathrm{F}}$  (Fig. S16 in the Supplementary Material~\cite{SM}, available in the published version).
As shown in Fig. \ref{Figure4}(e) and (f), when we set the energy at $E$ = $E_{\mathrm{F}}$ + 15 meV that crosses the $f$ flat bands, we can obtain $|\sigma_{xy}^A|$ being about $ 1000~\Omega^{-1}\,\text{cm}^{-1}$ and $|\alpha_{xy}^A|$ being about $4~\text{A}\,\text{m}^{-1}\,\text{K}^{-1}$ at 50~K, agree well with the experimental values.

From the perspective of intrinsic mechanisms, the AHE and the ANE arise from electron flows that are deflected by the BC under an electric field and a thermal gradient, respectively \cite{2015FundamentalsBehnia, Behnia2022TEP}. These two effects are connected via the Mott relation $\frac{\alpha^{\mathrm{A}}_{xy}}{T}= -\frac{\pi^2}{3} \frac{k^2_{\mathrm{B}}}{e} \frac{d\sigma^{\mathrm{A}}_{xy}}{d E}\Big|_{E = E_{\mathrm{F}}}$, where $k_{\mathrm{B}}$ is the Boltzmann constant and $e$ is the electron charge.
Apparently the ratio of $\alpha^{\mathrm{A}}_{xy}/\sigma^{\mathrm{A}}_{xy}$ for LaCo$_2$As$_2$ remains linear dependence on the temperature while that for CeCo$_2$As$_2$ deviates from the linearity above 30~K (Fig. 3(c)). To understand the deviation from linearity, we perform the Sommerfeld expansion, with the help of the Dirac model, to modify it as  
$\frac{\alpha^{\mathrm{A}}_{xy}}{\sigma^{\mathrm{A}}_{xy}}=\frac{eL_0T}{\mu[1+(\pi^2/3)(k_{\mathrm{B}}T/\mu)^2]}$, where $L_0$ is the Lorentz constant and $\mu$ is the chemical potential \cite{Lu2023Mott}. 
In the limit of $ k_{\mathrm{B}}T \ll \mu$, the second term of the denominator is negligible, and the ratio $\alpha^{\mathrm{A}}_{xy}/\sigma^{\mathrm{A}}_{xy}$ exhibits a linear temperature dependence, which is observed in LaCo$_2$As$_2$ and many other topological magnets \cite{crossoverbehaviorTukura2007,Pu2008GaMnAs,Mn3GeXu2020,Xu2020Mott,Xu2022Tb166}.
In contrast, the second term is not negligible for CeCo$_2$As$_2$, indicating a small $\mu$.
We fit the measured data with the above expression, as shown by the solid curves in Fig.~\ref{Figure3}(c), to extract the $\mu$ value corresponding to $T^{\ast}=\mu/k_{\mathrm{B}}=90$~K for CeCo$_2$As$_2$.
This value is comparable to $T_{\mathrm{coh}}=100$~K, indicating that the renormalized band due to the coherent Kondo hybridization is the source of the BC. On the other hand, its ultra-low Fermi temperature ($T_{\mathrm{F}}$) and $T_{\mathrm{coh}}$ are responsible for the breakdown of the Mott relation at elevated temperatures.
As comparison, the estimated $T^{\ast}$ for LaCo$_2$As$_2$ is 475~K, far above its $T_{\mathrm{C}}$.

We now demonstrate that the low $T_{\mathrm{F}}$ is crucial for the exceptional values of $S^{\mathrm{A}}_{xy}/T|_{T\rightarrow0}$ and $\tan\theta_{\mathrm{AN}}$ in CeCo$_2$As$_2$.
Given that the intrinsic $\sigma^{\mathrm{A}}_{xy}$ is independent of electronic scattering, a low $T_{\mathrm{F}}$ leads to an elevated value of  $\alpha_{xy}^{\mathrm{A}}/T|_{T\rightarrow0}$ in accordance with the Mott relation.
Because the second term $\sigma_{xx} S^{\mathrm{A}}_{xy}$ contributes over 90\% to $\alpha^{\mathrm{A}}_{xy}=\sigma^{\mathrm{A}}_{xy}S_{yy}+\sigma_{xx} S^{\mathrm{A}}_{xy}$ in CeCo$_2$As$_2$, a high value of $S^{\mathrm{A}}_{xy}/T|_{T\rightarrow0}$ is anticipated when $\sigma_{xx}$ is low.

On the other hand, by applying the Mott relation in the low-temperature limit for both $\alpha_{xy}^{\mathrm{A}}$ and $\alpha_{xx}$, we derive the expression $\tan\theta_{\mathrm{AN}}|_{T\rightarrow0}\simeq \left|\frac{d\sigma^{\mathrm{A}}_{xy}/d E}{d\sigma_{xx}/d E} \right|_{E = E_{\mathrm{F}}}$, which can be compared to the formula of $\tan\theta_{\mathrm{AH}}=\left|{\sigma^{\mathrm{A}}_{xy}}/{\sigma_{xx}}\right|$.
The large disparity between the two values ($\tan\theta_{\mathrm{AH}} = 11\%$ and $\tan\theta_{\mathrm{AN}} = 144\%$) indicates that $\sigma^{\mathrm{A}}_{xy}$ exhibits a markedly more sensitive response concerning energy variation than that of $\sigma_{xx}$.
This result is consistent with our calculations, which demonstrate a sharp energy dependence of \( \sigma^{\mathrm{A}}_{xy} \) near the Fermi level, as shown in Fig.~4(e).

\begin{figure}
	\includegraphics[width=1\linewidth]{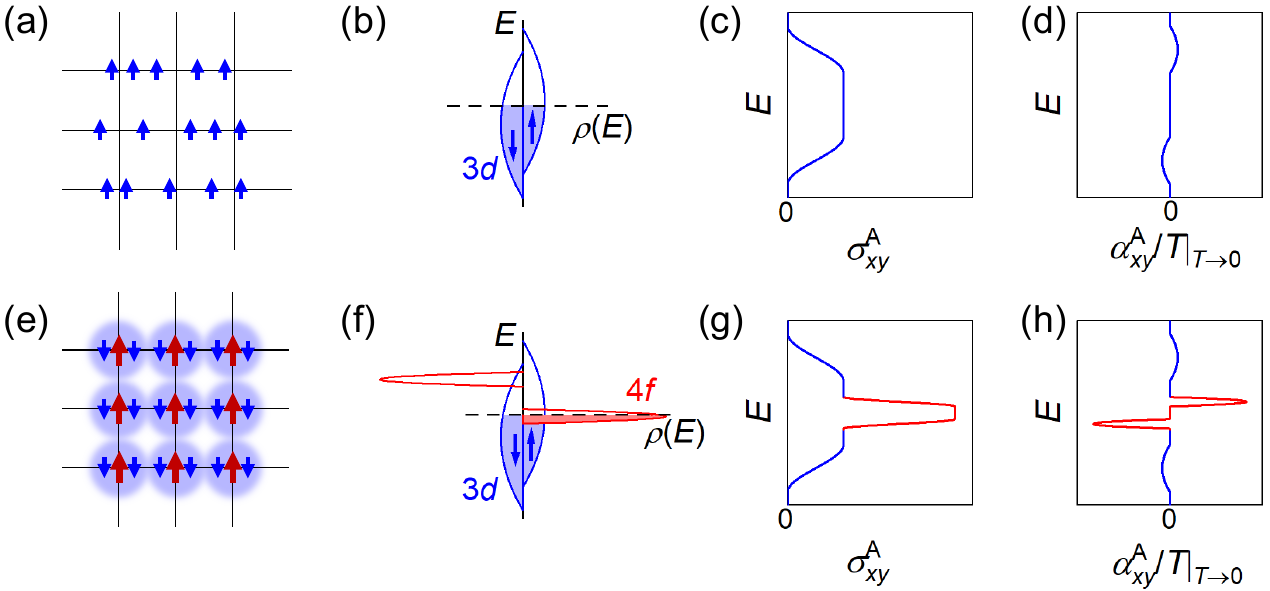} 
	\caption{Schematics of the configuration of spins, the DOS, $\sigma^{\mathrm{A}}_{xy}$, and low-temperature $\alpha^{\mathrm{A}}_{xy}$ divided by the temperature $\alpha_{xy}/T|_{T\rightarrow0}$ in (a)-(d) the itinerant ferromagnet, and (e)-(h) the FM Kondo lattice.} 
	\label{Figure5} 
\end{figure}

As a measurement of the BC at $E_{\mathrm{F}}$, a large ANE can be achieved only when the chemical potential is tuned in the vicinity of the BC `hot zones', which sometimes requires elaborate chemical substitution in topological magnets in practice \cite{Li2023Fe3Sn2,Liu2023Co3Sn2S2FeNi}. Our calculation shows that the hybridization in CeCo$_2$As$_2$ spreads over the BC to an energy range of approximately tens of meV (Fig. S16 in the Supplementary Material~\cite{SM}, available in the published version).
As the bandwidth is of the same magnitude, the BC is distributed over a significant portion of the flat band, leading to an enhanced ANE.

Recent studies revealed that an inversion-symmetry-breaking HF can evolve into a Weyl-Kondo semimetal state with strongly renormalized Weyl nodes residing at $E_{\mathrm{F}}$ \cite{lai_weylkondo_2018, WeylKS2019, Paschen2021topoKondo,Chen2022topocorre, Checkelsky2024flatband}.
Such non-magnetic, non-centrosymmetric Kondo semimetals manifest enhanced BC due to the Kondo flat band and spontaneous Hall effect at ultra-low temperatures \cite{dzsaber_giant_2021}.
On the other hand, our findings unveil significantly enhanced AHE and ANE in the FM Kondo lattice CeCo$_2$As$_2$, which serve as the fingerprint of the strong BC in topological magnets.
Figures \ref{Figure5}(e) to (h) illustrate the pathway of the formation of this topological FM Kondo lattice and the enhancement effect of AHE and ANE, in comparison to the counterpart with only $d$-electron magnetism in Figs. \ref{Figure5}(a) to (d).
As the Kondo band is generated from the hybridization between the magnetic $d$-electron framework and the embedded $f$ moments, the flat band naturally involves time-reversal symmetry breaking.

A considerable value of $\tan\theta_{\mathrm{AN}}$, and a large ratio of $S^{\mathrm{A}}_{xy}/T$ at low temperatures, can serve as indicators of low $E_{\mathrm{F}}$ and strong BC in the flat band. 
These observations directly point to a Kondo-pinning effect, similar to that proposed in the non-magnetic Weyl-Kondo semimetals \cite{Chen2022topocorre}.
Those results showcase prominent anomalous transport effects of a correlated topological Kondo magnet and also highlight a thermoelectric quantum effect of a flat-band system.

\vspace{10pt}

\section*{acknowledgments}
This work was supported by the National Natural Science Foundation of China (Grants No. 12225401, No.12141002, and No. 12074041), the National Key Research and Development Program of China (2021YFA1401902), Quantum Science and Technology --- National Science and Technology Major Project (Grant No. 2021ZD0302600), Innovation Program for Quantum Science and Technology (No. 2021ZD0302800), and the Fundamental Research Funds for the Central Universities (Grant. No. 2243300003). The work at Michigan State University was supported by the Beckman Young Investigator Program. The DFT+DMFT calculations were carried out with the high-performance computing cluster of Beijing Normal University in Zhuhai.

\bibliography{ref.bib}

\begin{thebibliography}{81}%
\makeatletter
\providecommand \@ifxundefined [1]{%
 \@ifx{#1\undefined}
}%
\providecommand \@ifnum [1]{%
 \ifnum #1\expandafter \@firstoftwo
 \else \expandafter \@secondoftwo
 \fi
}%
\providecommand \@ifx [1]{%
 \ifx #1\expandafter \@firstoftwo
 \else \expandafter \@secondoftwo
 \fi
}%
\providecommand \natexlab [1]{#1}%
\providecommand \enquote  [1]{``#1''}%
\providecommand \bibnamefont  [1]{#1}%
\providecommand \bibfnamefont [1]{#1}%
\providecommand \citenamefont [1]{#1}%
\providecommand \href@noop [0]{\@secondoftwo}%
\providecommand \href [0]{\begingroup \@sanitize@url \@href}%
\providecommand \@href[1]{\@@startlink{#1}\@@href}%
\providecommand \@@href[1]{\endgroup#1\@@endlink}%
\providecommand \@sanitize@url [0]{\catcode `\\12\catcode `\$12\catcode
  `\&12\catcode `\#12\catcode `\^12\catcode `\_12\catcode `\%12\relax}%
\providecommand \@@startlink[1]{}%
\providecommand \@@endlink[0]{}%
\providecommand \url  [0]{\begingroup\@sanitize@url \@url }%
\providecommand \@url [1]{\endgroup\@href {#1}{\urlprefix }}%
\providecommand \urlprefix  [0]{URL }%
\providecommand \Eprint [0]{\href }%
\providecommand \doibase [0]{https://doi.org/}%
\providecommand \selectlanguage [0]{\@gobble}%
\providecommand \bibinfo  [0]{\@secondoftwo}%
\providecommand \bibfield  [0]{\@secondoftwo}%
\providecommand \translation [1]{[#1]}%
\providecommand \BibitemOpen [0]{}%
\providecommand \bibitemStop [0]{}%
\providecommand \bibitemNoStop [0]{.\EOS\space}%
\providecommand \EOS [0]{\spacefactor3000\relax}%
\providecommand \BibitemShut  [1]{\csname bibitem#1\endcsname}%
\let\auto@bib@innerbib\@empty
\bibitem [{\citenamefont {Mizuguchi}\ and\ \citenamefont
  {Nakatsuji}(2019)}]{deviceNakatsuji2019}%
  \BibitemOpen
  \bibfield  {author} {\bibinfo {author} {\bibfnamefont {M.}~\bibnamefont
  {Mizuguchi}}\ and\ \bibinfo {author} {\bibfnamefont {S.}~\bibnamefont
  {Nakatsuji}},\ }\bibfield  {title} {\bibinfo {title} {Energy-harvesting
  materials based on the anomalous {N}ernst effect},\ }\href
  {https://doi.org/10.1080/14686996.2019.1585143} {\bibfield  {journal}
  {\bibinfo  {journal} {Sci. Technol. Adv. Mater.}\ }\textbf {\bibinfo {volume}
  {20}},\ \bibinfo {pages} {262} (\bibinfo {year} {2019})}\BibitemShut
  {NoStop}%
\bibitem [{\citenamefont {Fu}\ \emph {et~al.}(2020)\citenamefont {Fu},
  \citenamefont {Sun},\ and\ \citenamefont {Felser}}]{Fu2020TEP}%
  \BibitemOpen
  \bibfield  {author} {\bibinfo {author} {\bibfnamefont {C.}~\bibnamefont
  {Fu}}, \bibinfo {author} {\bibfnamefont {Y.}~\bibnamefont {Sun}},\ and\
  \bibinfo {author} {\bibfnamefont {C.}~\bibnamefont {Felser}},\ }\bibfield
  {title} {\bibinfo {title} {Topological thermoelectrics},\ }\href
  {https://doi.org/10.1063/5.0005481} {\bibfield  {journal} {\bibinfo
  {journal} {APL Mater.}\ }\textbf {\bibinfo {volume} {8}},\ \bibinfo {pages}
  {040913} (\bibinfo {year} {2020})}\BibitemShut {NoStop}%
\bibitem [{\citenamefont {Ramos}\ \emph {et~al.}(2014)\citenamefont {Ramos},
  \citenamefont {Aguirre}, \citenamefont {Anad\'on}, \citenamefont {Blasco},
  \citenamefont {Lucas}, \citenamefont {Uchida}, \citenamefont {Algarabel},
  \citenamefont {Morell\'on}, \citenamefont {Saitoh},\ and\ \citenamefont
  {Ibarra}}]{Ramos2014Fe3O4}%
  \BibitemOpen
  \bibfield  {author} {\bibinfo {author} {\bibfnamefont {R.}~\bibnamefont
  {Ramos}}, \bibinfo {author} {\bibfnamefont {M.~H.}\ \bibnamefont {Aguirre}},
  \bibinfo {author} {\bibfnamefont {A.}~\bibnamefont {Anad\'on}}, \bibinfo
  {author} {\bibfnamefont {J.}~\bibnamefont {Blasco}}, \bibinfo {author}
  {\bibfnamefont {I.}~\bibnamefont {Lucas}}, \bibinfo {author} {\bibfnamefont
  {K.}~\bibnamefont {Uchida}}, \bibinfo {author} {\bibfnamefont {P.~A.}\
  \bibnamefont {Algarabel}}, \bibinfo {author} {\bibfnamefont {L.}~\bibnamefont
  {Morell\'on}}, \bibinfo {author} {\bibfnamefont {E.}~\bibnamefont {Saitoh}},\
  and\ \bibinfo {author} {\bibfnamefont {M.~R.}\ \bibnamefont {Ibarra}},\
  }\bibfield  {title} {\bibinfo {title} {Anomalous {N}ernst effect of
  {F}e$_{3}${O}$_{4}$ single crystal},\ }\href
  {https://doi.org/10.1103/PhysRevB.90.054422} {\bibfield  {journal} {\bibinfo
  {journal} {Phys. Rev. B}\ }\textbf {\bibinfo {volume} {90}},\ \bibinfo
  {pages} {054422} (\bibinfo {year} {2014})}\BibitemShut {NoStop}%
\bibitem [{\citenamefont {Hasegawa}\ \emph {et~al.}(2015)\citenamefont
  {Hasegawa}, \citenamefont {Mizuguchi}, \citenamefont {Sakuraba},
  \citenamefont {Kamada}, \citenamefont {Kojima}, \citenamefont {Kubota},
  \citenamefont {Mizukami}, \citenamefont {Miyazaki},\ and\ \citenamefont
  {Takanashi}}]{Hasegawa2015Fe}%
  \BibitemOpen
  \bibfield  {author} {\bibinfo {author} {\bibfnamefont {K.}~\bibnamefont
  {Hasegawa}}, \bibinfo {author} {\bibfnamefont {M.}~\bibnamefont {Mizuguchi}},
  \bibinfo {author} {\bibfnamefont {Y.}~\bibnamefont {Sakuraba}}, \bibinfo
  {author} {\bibfnamefont {T.}~\bibnamefont {Kamada}}, \bibinfo {author}
  {\bibfnamefont {T.}~\bibnamefont {Kojima}}, \bibinfo {author} {\bibfnamefont
  {T.}~\bibnamefont {Kubota}}, \bibinfo {author} {\bibfnamefont
  {S.}~\bibnamefont {Mizukami}}, \bibinfo {author} {\bibfnamefont
  {T.}~\bibnamefont {Miyazaki}},\ and\ \bibinfo {author} {\bibfnamefont
  {K.}~\bibnamefont {Takanashi}},\ }\bibfield  {title} {\bibinfo {title}
  {Material dependence of anomalous {N}ernst effect in perpendicularly
  magnetized ordered-alloy thin films},\ }\href
  {https://doi.org/10.1063/1.4922901} {\bibfield  {journal} {\bibinfo
  {journal} {Appl. Phys. Lett.}\ }\textbf {\bibinfo {volume} {106}},\ \bibinfo
  {pages} {252405} (\bibinfo {year} {2015})}\BibitemShut {NoStop}%
\bibitem [{\citenamefont {Chuang}\ \emph {et~al.}(2017)\citenamefont {Chuang},
  \citenamefont {Su}, \citenamefont {Wu},\ and\ \citenamefont
  {Huang}}]{Chuang2017FeCoNi}%
  \BibitemOpen
  \bibfield  {author} {\bibinfo {author} {\bibfnamefont {T.~C.}\ \bibnamefont
  {Chuang}}, \bibinfo {author} {\bibfnamefont {P.~L.}\ \bibnamefont {Su}},
  \bibinfo {author} {\bibfnamefont {P.~H.}\ \bibnamefont {Wu}},\ and\ \bibinfo
  {author} {\bibfnamefont {S.~Y.}\ \bibnamefont {Huang}},\ }\bibfield  {title}
  {\bibinfo {title} {Enhancement of the anomalous {N}ernst effect in
  ferromagnetic thin films},\ }\href
  {https://doi.org/10.1103/PhysRevB.96.174406} {\bibfield  {journal} {\bibinfo
  {journal} {Phys. Rev. B}\ }\textbf {\bibinfo {volume} {96}},\ \bibinfo
  {pages} {174406} (\bibinfo {year} {2017})}\BibitemShut {NoStop}%
\bibitem [{\citenamefont {Xiao}\ \emph {et~al.}(2006)\citenamefont {Xiao},
  \citenamefont {Yao}, \citenamefont {Fang},\ and\ \citenamefont
  {Niu}}]{BerryXiao2006}%
  \BibitemOpen
  \bibfield  {author} {\bibinfo {author} {\bibfnamefont {D.}~\bibnamefont
  {Xiao}}, \bibinfo {author} {\bibfnamefont {Y.}~\bibnamefont {Yao}}, \bibinfo
  {author} {\bibfnamefont {Z.}~\bibnamefont {Fang}},\ and\ \bibinfo {author}
  {\bibfnamefont {Q.}~\bibnamefont {Niu}},\ }\bibfield  {title} {\bibinfo
  {title} {Berry-phase effect in anomalous thermoelectric transport},\ }\href
  {https://doi.org/10.1103/PhysRevLett.97.026603} {\bibfield  {journal}
  {\bibinfo  {journal} {Phys. Rev. Lett.}\ }\textbf {\bibinfo {volume} {97}},\
  \bibinfo {pages} {026603} (\bibinfo {year} {2006})}\BibitemShut {NoStop}%
\bibitem [{\citenamefont {Ikhlas}\ \emph {et~al.}(2017)\citenamefont {Ikhlas},
  \citenamefont {Tomita}, \citenamefont {Koretsune}, \citenamefont {Suzuki},
  \citenamefont {Nishio-Hamane}, \citenamefont {Arita}, \citenamefont {Otani},\
  and\ \citenamefont {Nakatsuji}}]{Mn3SnIkhlas2017}%
  \BibitemOpen
  \bibfield  {author} {\bibinfo {author} {\bibfnamefont {M.}~\bibnamefont
  {Ikhlas}}, \bibinfo {author} {\bibfnamefont {T.}~\bibnamefont {Tomita}},
  \bibinfo {author} {\bibfnamefont {T.}~\bibnamefont {Koretsune}}, \bibinfo
  {author} {\bibfnamefont {M.-T.}\ \bibnamefont {Suzuki}}, \bibinfo {author}
  {\bibfnamefont {D.}~\bibnamefont {Nishio-Hamane}}, \bibinfo {author}
  {\bibfnamefont {R.}~\bibnamefont {Arita}}, \bibinfo {author} {\bibfnamefont
  {Y.}~\bibnamefont {Otani}},\ and\ \bibinfo {author} {\bibfnamefont
  {S.}~\bibnamefont {Nakatsuji}},\ }\bibfield  {title} {\bibinfo {title} {Large
  anomalous {N}ernst effect at room temperature in a chiral antiferromagnet},\
  }\href {https://doi.org/10.1038/nphys4181} {\bibfield  {journal} {\bibinfo
  {journal} {Nat. Phys.}\ }\textbf {\bibinfo {volume} {13}},\ \bibinfo {pages}
  {1085} (\bibinfo {year} {2017})}\BibitemShut {NoStop}%
\bibitem [{\citenamefont {Sakai}\ \emph {et~al.}(2018)\citenamefont {Sakai},
  \citenamefont {Mizuta}, \citenamefont {Nugroho}, \citenamefont {Sihombing},
  \citenamefont {Koretsune}, \citenamefont {Suzuki}, \citenamefont {Takemori},
  \citenamefont {Ishii}, \citenamefont {Nishio-Hamane}, \citenamefont {Arita},
  \citenamefont {Goswami},\ and\ \citenamefont {Nakatsuji}}]{Co2MnGaSakai2018}%
  \BibitemOpen
  \bibfield  {author} {\bibinfo {author} {\bibfnamefont {A.}~\bibnamefont
  {Sakai}}, \bibinfo {author} {\bibfnamefont {Y.~P.}\ \bibnamefont {Mizuta}},
  \bibinfo {author} {\bibfnamefont {A.~A.}\ \bibnamefont {Nugroho}}, \bibinfo
  {author} {\bibfnamefont {R.}~\bibnamefont {Sihombing}}, \bibinfo {author}
  {\bibfnamefont {T.}~\bibnamefont {Koretsune}}, \bibinfo {author}
  {\bibfnamefont {M.-T.}\ \bibnamefont {Suzuki}}, \bibinfo {author}
  {\bibfnamefont {N.}~\bibnamefont {Takemori}}, \bibinfo {author}
  {\bibfnamefont {R.}~\bibnamefont {Ishii}}, \bibinfo {author} {\bibfnamefont
  {D.}~\bibnamefont {Nishio-Hamane}}, \bibinfo {author} {\bibfnamefont
  {R.}~\bibnamefont {Arita}}, \bibinfo {author} {\bibfnamefont
  {P.}~\bibnamefont {Goswami}},\ and\ \bibinfo {author} {\bibfnamefont
  {S.}~\bibnamefont {Nakatsuji}},\ }\bibfield  {title} {\bibinfo {title} {Giant
  anomalous {N}ernst effect and quantum-critical scaling in a ferromagnetic
  semimetal},\ }\href {https://doi.org/10.1038/s41567-018-0225-6} {\bibfield
  {journal} {\bibinfo  {journal} {Nat. Phys.}\ }\textbf {\bibinfo {volume}
  {14}},\ \bibinfo {pages} {1119} (\bibinfo {year} {2018})}\BibitemShut
  {NoStop}%
\bibitem [{\citenamefont {Guin}\ \emph
  {et~al.}(2019{\natexlab{a}})\citenamefont {Guin}, \citenamefont {Vir},
  \citenamefont {Zhang}, \citenamefont {Kumar}, \citenamefont {Watzman},
  \citenamefont {Fu}, \citenamefont {Liu}, \citenamefont {Manna}, \citenamefont
  {Schnelle}, \citenamefont {Gooth}, \citenamefont {Shekhar}, \citenamefont
  {Sun},\ and\ \citenamefont {Felser}}]{Co3Sn2S2Guin2018}%
  \BibitemOpen
  \bibfield  {author} {\bibinfo {author} {\bibfnamefont {S.~N.}\ \bibnamefont
  {Guin}}, \bibinfo {author} {\bibfnamefont {P.}~\bibnamefont {Vir}}, \bibinfo
  {author} {\bibfnamefont {Y.}~\bibnamefont {Zhang}}, \bibinfo {author}
  {\bibfnamefont {N.}~\bibnamefont {Kumar}}, \bibinfo {author} {\bibfnamefont
  {S.~J.}\ \bibnamefont {Watzman}}, \bibinfo {author} {\bibfnamefont
  {C.}~\bibnamefont {Fu}}, \bibinfo {author} {\bibfnamefont {E.}~\bibnamefont
  {Liu}}, \bibinfo {author} {\bibfnamefont {K.}~\bibnamefont {Manna}}, \bibinfo
  {author} {\bibfnamefont {W.}~\bibnamefont {Schnelle}}, \bibinfo {author}
  {\bibfnamefont {J.}~\bibnamefont {Gooth}}, \bibinfo {author} {\bibfnamefont
  {C.}~\bibnamefont {Shekhar}}, \bibinfo {author} {\bibfnamefont
  {Y.}~\bibnamefont {Sun}},\ and\ \bibinfo {author} {\bibfnamefont
  {C.}~\bibnamefont {Felser}},\ }\bibfield  {title} {\bibinfo {title}
  {Zero-field {N}ernst effect in a ferromagnetic kagome-lattice
  {W}eyl-semimetal {C}o$_3${S}n$_2${S}$_2$},\ }\href
  {https://doi.org/https://doi.org/10.1002/adma.201806622} {\bibfield
  {journal} {\bibinfo  {journal} {Adv. Mater.}\ }\textbf {\bibinfo {volume}
  {31}},\ \bibinfo {pages} {1806622} (\bibinfo {year}
  {2019}{\natexlab{a}})}\BibitemShut {NoStop}%
\bibitem [{\citenamefont {Xu}\ \emph {et~al.}(2019)\citenamefont {Xu},
  \citenamefont {Phelan},\ and\ \citenamefont {Chien}}]{Fe3GeTe2Xu2019}%
  \BibitemOpen
  \bibfield  {author} {\bibinfo {author} {\bibfnamefont {J.}~\bibnamefont
  {Xu}}, \bibinfo {author} {\bibfnamefont {W.~A.}\ \bibnamefont {Phelan}},\
  and\ \bibinfo {author} {\bibfnamefont {C.-L.}\ \bibnamefont {Chien}},\
  }\bibfield  {title} {\bibinfo {title} {Large anomalous {N}ernst effect in a
  van der {W}aals ferromagnet {F}e$_3${G}e{T}e$_2$},\ }\href
  {https://doi.org/10.1021/acs.nanolett.9b03739} {\bibfield  {journal}
  {\bibinfo  {journal} {Nano Lett.}\ }\textbf {\bibinfo {volume} {19}},\
  \bibinfo {pages} {8250} (\bibinfo {year} {2019})}\BibitemShut {NoStop}%
\bibitem [{\citenamefont {Sakai}\ \emph {et~al.}(2020)\citenamefont {Sakai},
  \citenamefont {Minami}, \citenamefont {Koretsune}, \citenamefont {Chen},
  \citenamefont {Higo}, \citenamefont {Wang}, \citenamefont {Nomoto},
  \citenamefont {Hirayama}, \citenamefont {Miwa}, \citenamefont
  {Nishio-Hamane}, \citenamefont {Ishii}, \citenamefont {Arita},\ and\
  \citenamefont {Nakatsuji}}]{Fe3GaSakai2020}%
  \BibitemOpen
  \bibfield  {author} {\bibinfo {author} {\bibfnamefont {A.}~\bibnamefont
  {Sakai}}, \bibinfo {author} {\bibfnamefont {S.}~\bibnamefont {Minami}},
  \bibinfo {author} {\bibfnamefont {T.}~\bibnamefont {Koretsune}}, \bibinfo
  {author} {\bibfnamefont {T.}~\bibnamefont {Chen}}, \bibinfo {author}
  {\bibfnamefont {T.}~\bibnamefont {Higo}}, \bibinfo {author} {\bibfnamefont
  {Y.}~\bibnamefont {Wang}}, \bibinfo {author} {\bibfnamefont {T.}~\bibnamefont
  {Nomoto}}, \bibinfo {author} {\bibfnamefont {M.}~\bibnamefont {Hirayama}},
  \bibinfo {author} {\bibfnamefont {S.}~\bibnamefont {Miwa}}, \bibinfo {author}
  {\bibfnamefont {D.}~\bibnamefont {Nishio-Hamane}}, \bibinfo {author}
  {\bibfnamefont {F.}~\bibnamefont {Ishii}}, \bibinfo {author} {\bibfnamefont
  {R.}~\bibnamefont {Arita}},\ and\ \bibinfo {author} {\bibfnamefont
  {S.}~\bibnamefont {Nakatsuji}},\ }\bibfield  {title} {\bibinfo {title}
  {Iron-based binary ferromagnets for transverse thermoelectric conversion},\
  }\href {https://doi.org/10.1038/s41586-020-2230-z} {\bibfield  {journal}
  {\bibinfo  {journal} {Nature}\ }\textbf {\bibinfo {volume} {581}},\ \bibinfo
  {pages} {53} (\bibinfo {year} {2020})}\BibitemShut {NoStop}%
\bibitem [{\citenamefont {Xu}\ \emph {et~al.}(2020{\natexlab{a}})\citenamefont
  {Xu}, \citenamefont {Li}, \citenamefont {Lu}, \citenamefont {Collignon},
  \citenamefont {Fu}, \citenamefont {Koo}, \citenamefont {Fauqué},
  \citenamefont {Yan}, \citenamefont {Zhu},\ and\ \citenamefont
  {Behnia}}]{Mn3GeXu2020}%
  \BibitemOpen
  \bibfield  {author} {\bibinfo {author} {\bibfnamefont {L.}~\bibnamefont
  {Xu}}, \bibinfo {author} {\bibfnamefont {X.}~\bibnamefont {Li}}, \bibinfo
  {author} {\bibfnamefont {X.}~\bibnamefont {Lu}}, \bibinfo {author}
  {\bibfnamefont {C.}~\bibnamefont {Collignon}}, \bibinfo {author}
  {\bibfnamefont {H.}~\bibnamefont {Fu}}, \bibinfo {author} {\bibfnamefont
  {J.}~\bibnamefont {Koo}}, \bibinfo {author} {\bibfnamefont {B.}~\bibnamefont
  {Fauqué}}, \bibinfo {author} {\bibfnamefont {B.}~\bibnamefont {Yan}},
  \bibinfo {author} {\bibfnamefont {Z.}~\bibnamefont {Zhu}},\ and\ \bibinfo
  {author} {\bibfnamefont {K.}~\bibnamefont {Behnia}},\ }\bibfield  {title}
  {\bibinfo {title} {Finite-temperature violation of the anomalous transverse
  {W}iedemann-{F}ranz law},\ }\href {https://doi.org/10.1126/sciadv.aaz3522}
  {\bibfield  {journal} {\bibinfo  {journal} {Sci. Adv.}\ }\textbf {\bibinfo
  {volume} {6}},\ \bibinfo {pages} {eaaz3522} (\bibinfo {year}
  {2020}{\natexlab{a}})}\BibitemShut {NoStop}%
\bibitem [{\citenamefont {Zhang}\ \emph {et~al.}(2021)\citenamefont {Zhang},
  \citenamefont {Xu},\ and\ \citenamefont {Ke}}]{Fe3Sn2Zhang2021}%
  \BibitemOpen
  \bibfield  {author} {\bibinfo {author} {\bibfnamefont {H.}~\bibnamefont
  {Zhang}}, \bibinfo {author} {\bibfnamefont {C.~Q.}\ \bibnamefont {Xu}},\ and\
  \bibinfo {author} {\bibfnamefont {X.}~\bibnamefont {Ke}},\ }\bibfield
  {title} {\bibinfo {title} {Topological {N}ernst effect, anomalous {N}ernst
  effect, and anomalous thermal hall effect in the {D}irac semimetal
  {F}e$_3${S}n$_2$},\ }\href {https://doi.org/10.1103/PhysRevB.103.L201101}
  {\bibfield  {journal} {\bibinfo  {journal} {Phys. Rev. B}\ }\textbf {\bibinfo
  {volume} {103}},\ \bibinfo {pages} {L201101} (\bibinfo {year}
  {2021})}\BibitemShut {NoStop}%
\bibitem [{\citenamefont {Asaba}\ \emph {et~al.}(2021)\citenamefont {Asaba},
  \citenamefont {Ivanov}, \citenamefont {Thomas}, \citenamefont {Savrasov},
  \citenamefont {Thompson}, \citenamefont {Bauer},\ and\ \citenamefont
  {Ronning}}]{UCoRuAlAsaba2021}%
  \BibitemOpen
  \bibfield  {author} {\bibinfo {author} {\bibfnamefont {T.}~\bibnamefont
  {Asaba}}, \bibinfo {author} {\bibfnamefont {V.}~\bibnamefont {Ivanov}},
  \bibinfo {author} {\bibfnamefont {S.~M.}\ \bibnamefont {Thomas}}, \bibinfo
  {author} {\bibfnamefont {S.~Y.}\ \bibnamefont {Savrasov}}, \bibinfo {author}
  {\bibfnamefont {J.~D.}\ \bibnamefont {Thompson}}, \bibinfo {author}
  {\bibfnamefont {E.~D.}\ \bibnamefont {Bauer}},\ and\ \bibinfo {author}
  {\bibfnamefont {F.}~\bibnamefont {Ronning}},\ }\bibfield  {title} {\bibinfo
  {title} {Colossal anomalous {N}ernst effect in a correlated
  noncentrosymmetric kagome ferromagnet},\ }\href
  {https://doi.org/10.1126/sciadv.abf1467} {\bibfield  {journal} {\bibinfo
  {journal} {Sci. Adv.}\ }\textbf {\bibinfo {volume} {7}},\ \bibinfo {pages}
  {eabf1467} (\bibinfo {year} {2021})}\BibitemShut {NoStop}%
\bibitem [{\citenamefont {Pan}\ \emph {et~al.}(2022)\citenamefont {Pan},
  \citenamefont {Le}, \citenamefont {He}, \citenamefont {Watzman},
  \citenamefont {Yao}, \citenamefont {Gooth}, \citenamefont {Heremans},
  \citenamefont {Sun},\ and\ \citenamefont {Felser}}]{YbMnBi2Pan2022}%
  \BibitemOpen
  \bibfield  {author} {\bibinfo {author} {\bibfnamefont {Y.}~\bibnamefont
  {Pan}}, \bibinfo {author} {\bibfnamefont {C.}~\bibnamefont {Le}}, \bibinfo
  {author} {\bibfnamefont {B.}~\bibnamefont {He}}, \bibinfo {author}
  {\bibfnamefont {S.~J.}\ \bibnamefont {Watzman}}, \bibinfo {author}
  {\bibfnamefont {M.}~\bibnamefont {Yao}}, \bibinfo {author} {\bibfnamefont
  {J.}~\bibnamefont {Gooth}}, \bibinfo {author} {\bibfnamefont {J.~P.}\
  \bibnamefont {Heremans}}, \bibinfo {author} {\bibfnamefont {Y.}~\bibnamefont
  {Sun}},\ and\ \bibinfo {author} {\bibfnamefont {C.}~\bibnamefont {Felser}},\
  }\bibfield  {title} {\bibinfo {title} {Giant anomalous {N}ernst signal in the
  antiferromagnet {Y}b{M}n{B}i$_2$},\ }\href
  {https://doi.org/10.1038/s41563-021-01149-2} {\bibfield  {journal} {\bibinfo
  {journal} {Nat. Mater.}\ }\textbf {\bibinfo {volume} {21}},\ \bibinfo {pages}
  {203} (\bibinfo {year} {2022})}\BibitemShut {NoStop}%
\bibitem [{\citenamefont {Chen}\ \emph
  {et~al.}(2022{\natexlab{a}})\citenamefont {Chen}, \citenamefont {Minami},
  \citenamefont {Sakai}, \citenamefont {Wang}, \citenamefont {Feng},
  \citenamefont {Nomoto}, \citenamefont {Hirayama}, \citenamefont {Ishii},
  \citenamefont {Koretsune}, \citenamefont {Arita},\ and\ \citenamefont
  {Nakatsuji}}]{Fe3SnChen2022}%
  \BibitemOpen
  \bibfield  {author} {\bibinfo {author} {\bibfnamefont {T.}~\bibnamefont
  {Chen}}, \bibinfo {author} {\bibfnamefont {S.}~\bibnamefont {Minami}},
  \bibinfo {author} {\bibfnamefont {A.}~\bibnamefont {Sakai}}, \bibinfo
  {author} {\bibfnamefont {Y.}~\bibnamefont {Wang}}, \bibinfo {author}
  {\bibfnamefont {Z.}~\bibnamefont {Feng}}, \bibinfo {author} {\bibfnamefont
  {T.}~\bibnamefont {Nomoto}}, \bibinfo {author} {\bibfnamefont
  {M.}~\bibnamefont {Hirayama}}, \bibinfo {author} {\bibfnamefont
  {R.}~\bibnamefont {Ishii}}, \bibinfo {author} {\bibfnamefont
  {T.}~\bibnamefont {Koretsune}}, \bibinfo {author} {\bibfnamefont
  {R.}~\bibnamefont {Arita}},\ and\ \bibinfo {author} {\bibfnamefont
  {S.}~\bibnamefont {Nakatsuji}},\ }\bibfield  {title} {\bibinfo {title} {Large
  anomalous {N}ernst effect and nodal plane in an iron-based kagome
  ferromagnet},\ }\href {https://doi.org/10.1126/sciadv.abk1480} {\bibfield
  {journal} {\bibinfo  {journal} {Sci. Adv.}\ }\textbf {\bibinfo {volume}
  {8}},\ \bibinfo {pages} {eabk1480} (\bibinfo {year}
  {2022}{\natexlab{a}})}\BibitemShut {NoStop}%
\bibitem [{\citenamefont {Li}\ \emph {et~al.}(2024)\citenamefont {Li},
  \citenamefont {Guan}, \citenamefont {Chi}, \citenamefont {Li}, \citenamefont
  {Lin}, \citenamefont {Xu},\ and\ \citenamefont {Jia}}]{CeCrGe3Li2024}%
  \BibitemOpen
  \bibfield  {author} {\bibinfo {author} {\bibfnamefont {L.}~\bibnamefont
  {Li}}, \bibinfo {author} {\bibfnamefont {S.}~\bibnamefont {Guan}}, \bibinfo
  {author} {\bibfnamefont {S.}~\bibnamefont {Chi}}, \bibinfo {author}
  {\bibfnamefont {J.}~\bibnamefont {Li}}, \bibinfo {author} {\bibfnamefont
  {X.}~\bibnamefont {Lin}}, \bibinfo {author} {\bibfnamefont {G.}~\bibnamefont
  {Xu}},\ and\ \bibinfo {author} {\bibfnamefont {S.}~\bibnamefont {Jia}},\
  }\href@noop {} {\bibinfo {title} {Giant anomalous {H}all and {N}ernst effects
  in a heavy fermion ferromagnet}} (\bibinfo {year} {2024}),\ \Eprint
  {https://arxiv.org/abs/2401.17624} {arXiv:2401.17624} \BibitemShut {NoStop}%
\bibitem [{\citenamefont {Noky}\ \emph
  {et~al.}(2018{\natexlab{a}})\citenamefont {Noky}, \citenamefont {Gayles},
  \citenamefont {Felser},\ and\ \citenamefont {Sun}}]{Noky2018Ti2MnAl}%
  \BibitemOpen
  \bibfield  {author} {\bibinfo {author} {\bibfnamefont {J.}~\bibnamefont
  {Noky}}, \bibinfo {author} {\bibfnamefont {J.}~\bibnamefont {Gayles}},
  \bibinfo {author} {\bibfnamefont {C.}~\bibnamefont {Felser}},\ and\ \bibinfo
  {author} {\bibfnamefont {Y.}~\bibnamefont {Sun}},\ }\bibfield  {title}
  {\bibinfo {title} {Strong anomalous {N}ernst effect in collinear magnetic
  {W}eyl semimetals without net magnetic moments},\ }\href
  {https://doi.org/10.1103/PhysRevB.97.220405} {\bibfield  {journal} {\bibinfo
  {journal} {Phys. Rev. B}\ }\textbf {\bibinfo {volume} {97}},\ \bibinfo
  {pages} {220405} (\bibinfo {year} {2018}{\natexlab{a}})}\BibitemShut
  {NoStop}%
\bibitem [{\citenamefont {Noky}\ \emph {et~al.}(2019)\citenamefont {Noky},
  \citenamefont {Xu}, \citenamefont {Felser},\ and\ \citenamefont
  {Sun}}]{Noky2019Fe2MnX}%
  \BibitemOpen
  \bibfield  {author} {\bibinfo {author} {\bibfnamefont {J.}~\bibnamefont
  {Noky}}, \bibinfo {author} {\bibfnamefont {Q.}~\bibnamefont {Xu}}, \bibinfo
  {author} {\bibfnamefont {C.}~\bibnamefont {Felser}},\ and\ \bibinfo {author}
  {\bibfnamefont {Y.}~\bibnamefont {Sun}},\ }\bibfield  {title} {\bibinfo
  {title} {Large anomalous {H}all and {N}ernst effects from nodal line symmetry
  breaking in {F}e$_{2}\mathrm{{M}n}{X}$ (${X}$ = {P}, {A}s, {S}b)},\ }\href
  {https://doi.org/10.1103/PhysRevB.99.165117} {\bibfield  {journal} {\bibinfo
  {journal} {Phys. Rev. B}\ }\textbf {\bibinfo {volume} {99}},\ \bibinfo
  {pages} {165117} (\bibinfo {year} {2019})}\BibitemShut {NoStop}%
\bibitem [{\citenamefont {Noky}\ \emph {et~al.}(2020)\citenamefont {Noky},
  \citenamefont {Zhang}, \citenamefont {Gooth}, \citenamefont {Felser},\ and\
  \citenamefont {Sun}}]{Noky2020Heusler}%
  \BibitemOpen
  \bibfield  {author} {\bibinfo {author} {\bibfnamefont {J.}~\bibnamefont
  {Noky}}, \bibinfo {author} {\bibfnamefont {Y.}~\bibnamefont {Zhang}},
  \bibinfo {author} {\bibfnamefont {J.}~\bibnamefont {Gooth}}, \bibinfo
  {author} {\bibfnamefont {C.}~\bibnamefont {Felser}},\ and\ \bibinfo {author}
  {\bibfnamefont {Y.}~\bibnamefont {Sun}},\ }\bibfield  {title} {\bibinfo
  {title} {Giant anomalous {H}all and {N}ernst effect in magnetic cubic
  {H}eusler compounds},\ }\href {https://doi.org/10.1038/s41524-020-0342-5}
  {\bibfield  {journal} {\bibinfo  {journal} {Npj Comput. Mater.}\ }\textbf
  {\bibinfo {volume} {6}},\ \bibinfo {pages} {77} (\bibinfo {year}
  {2020})}\BibitemShut {NoStop}%
\bibitem [{\citenamefont {Li}\ \emph {et~al.}(2023)\citenamefont {Li},
  \citenamefont {Zhou}, \citenamefont {Li}, \citenamefont {Qiao}, \citenamefont
  {Jiang}, \citenamefont {Chen}, \citenamefont {Li}, \citenamefont {Tao},\ and\
  \citenamefont {Xu}}]{Li2023Fe3Sn2}%
  \BibitemOpen
  \bibfield  {author} {\bibinfo {author} {\bibfnamefont {Y.}~\bibnamefont
  {Li}}, \bibinfo {author} {\bibfnamefont {J.}~\bibnamefont {Zhou}}, \bibinfo
  {author} {\bibfnamefont {M.}~\bibnamefont {Li}}, \bibinfo {author}
  {\bibfnamefont {L.}~\bibnamefont {Qiao}}, \bibinfo {author} {\bibfnamefont
  {C.}~\bibnamefont {Jiang}}, \bibinfo {author} {\bibfnamefont
  {Q.}~\bibnamefont {Chen}}, \bibinfo {author} {\bibfnamefont {Y.}~\bibnamefont
  {Li}}, \bibinfo {author} {\bibfnamefont {Q.}~\bibnamefont {Tao}},\ and\
  \bibinfo {author} {\bibfnamefont {Z.-A.}\ \bibnamefont {Xu}},\ }\bibfield
  {title} {\bibinfo {title} {Enhanced anomalous {N}ernst effect by tuning the
  chemical potential in the topological kagome ferromagnet
  {F}e$_{3}${S}n$_{2}$},\ }\href
  {https://doi.org/10.1103/PhysRevApplied.19.014026} {\bibfield  {journal}
  {\bibinfo  {journal} {Phys. Rev. Appl.}\ }\textbf {\bibinfo {volume} {19}},\
  \bibinfo {pages} {014026} (\bibinfo {year} {2023})}\BibitemShut {NoStop}%
\bibitem [{\citenamefont {Liu}\ \emph {et~al.}(2023)\citenamefont {Liu},
  \citenamefont {Ding}, \citenamefont {Xu}, \citenamefont {Li}, \citenamefont
  {Behnia},\ and\ \citenamefont {Zhu}}]{Liu2023Co3Sn2S2FeNi}%
  \BibitemOpen
  \bibfield  {author} {\bibinfo {author} {\bibfnamefont {J.}~\bibnamefont
  {Liu}}, \bibinfo {author} {\bibfnamefont {L.}~\bibnamefont {Ding}}, \bibinfo
  {author} {\bibfnamefont {L.}~\bibnamefont {Xu}}, \bibinfo {author}
  {\bibfnamefont {X.}~\bibnamefont {Li}}, \bibinfo {author} {\bibfnamefont
  {K.}~\bibnamefont {Behnia}},\ and\ \bibinfo {author} {\bibfnamefont
  {Z.}~\bibnamefont {Zhu}},\ }\bibfield  {title} {\bibinfo {title} {Tuning the
  anomalous {N}ernst and {H}all effects with shifting the chemical potential in
  {F}e-doped and {N}i-doped {C}o$_3${S}n$_2${S}$_{2}$},\ }\href
  {https://doi.org/10.1088/1361-648X/acdcd9} {\bibfield  {journal} {\bibinfo
  {journal} {J. Phys.:Condens. Matter}\ }\textbf {\bibinfo {volume} {35}},\
  \bibinfo {pages} {375501} (\bibinfo {year} {2023})}\BibitemShut {NoStop}%
\bibitem [{\citenamefont {Thompson}\ \emph {et~al.}(2014)\citenamefont
  {Thompson}, \citenamefont {Tan}, \citenamefont {Kovnir}, \citenamefont
  {Garlea}, \citenamefont {Gippius}, \citenamefont {Yaroslavtsev},
  \citenamefont {Menushenkov}, \citenamefont {Chernikov}, \citenamefont
  {B{\"u}ttgen}, \citenamefont {Kr{\"a}tschmer}, \citenamefont {Zubavichus},\
  and\ \citenamefont {Shatruk}}]{CCAThompson2014}%
  \BibitemOpen
  \bibfield  {author} {\bibinfo {author} {\bibfnamefont {C.~M.}\ \bibnamefont
  {Thompson}}, \bibinfo {author} {\bibfnamefont {X.}~\bibnamefont {Tan}},
  \bibinfo {author} {\bibfnamefont {K.}~\bibnamefont {Kovnir}}, \bibinfo
  {author} {\bibfnamefont {V.~O.}\ \bibnamefont {Garlea}}, \bibinfo {author}
  {\bibfnamefont {A.~A.}\ \bibnamefont {Gippius}}, \bibinfo {author}
  {\bibfnamefont {A.~A.}\ \bibnamefont {Yaroslavtsev}}, \bibinfo {author}
  {\bibfnamefont {A.~P.}\ \bibnamefont {Menushenkov}}, \bibinfo {author}
  {\bibfnamefont {R.~V.}\ \bibnamefont {Chernikov}}, \bibinfo {author}
  {\bibfnamefont {N.}~\bibnamefont {B{\"u}ttgen}}, \bibinfo {author}
  {\bibfnamefont {W.}~\bibnamefont {Kr{\"a}tschmer}}, \bibinfo {author}
  {\bibfnamefont {Y.~V.}\ \bibnamefont {Zubavichus}},\ and\ \bibinfo {author}
  {\bibfnamefont {M.}~\bibnamefont {Shatruk}},\ }\bibfield  {title} {\bibinfo
  {title} {Synthesis, structures, and magnetic properties of rare-earth cobalt
  arsenides, {R}{C}o$_2${A}s$_2$ ({R} = {L}a, {C}e, {P}r, {N}d)},\ }\href
  {https://doi.org/10.1021/cm501522v} {\bibfield  {journal} {\bibinfo
  {journal} {Chem. Mater.}\ }\textbf {\bibinfo {volume} {26}},\ \bibinfo
  {pages} {3825} (\bibinfo {year} {2014})}\BibitemShut {NoStop}%
\bibitem [{\citenamefont {Tan}\ \emph {et~al.}(2018)\citenamefont {Tan},
  \citenamefont {Tener},\ and\ \citenamefont {Shatruk}}]{CCATan2018}%
  \BibitemOpen
  \bibfield  {author} {\bibinfo {author} {\bibfnamefont {X.}~\bibnamefont
  {Tan}}, \bibinfo {author} {\bibfnamefont {Z.~P.}\ \bibnamefont {Tener}},\
  and\ \bibinfo {author} {\bibfnamefont {M.}~\bibnamefont {Shatruk}},\
  }\bibfield  {title} {\bibinfo {title} {Correlating itinerant magnetism in
  {R}{C}o$_2${P}n$_2$ pnictides ({R} = {L}a, {C}e, {P}r, {N}d, {E}u, {C}a; {P}n
  = {P}, {A}s) to their crystal and electronic structures},\ }\href
  {https://doi.org/10.1021/acs.accounts.7b00533} {\bibfield  {journal}
  {\bibinfo  {journal} {Acc. Chem. Res.}\ }\textbf {\bibinfo {volume} {51}},\
  \bibinfo {pages} {230} (\bibinfo {year} {2018})}\BibitemShut {NoStop}%
\bibitem [{\citenamefont {Huang}\ \emph {et~al.}(2023)\citenamefont {Huang},
  \citenamefont {Zheng}, \citenamefont {Liu}, \citenamefont {Xu}, \citenamefont
  {Wu}, \citenamefont {Dong}, \citenamefont {Wang}, \citenamefont {Yin},\ and\
  \citenamefont {Jia}}]{Huang2023anomalous}%
  \BibitemOpen
  \bibfield  {author} {\bibinfo {author} {\bibfnamefont {Y.-Q.}\ \bibnamefont
  {Huang}}, \bibinfo {author} {\bibfnamefont {P.-Y.}\ \bibnamefont {Zheng}},
  \bibinfo {author} {\bibfnamefont {R.}~\bibnamefont {Liu}}, \bibinfo {author}
  {\bibfnamefont {X.-T.}\ \bibnamefont {Xu}}, \bibinfo {author} {\bibfnamefont
  {Z.-Y.}\ \bibnamefont {Wu}}, \bibinfo {author} {\bibfnamefont
  {C.}~\bibnamefont {Dong}}, \bibinfo {author} {\bibfnamefont {J.-F.}\
  \bibnamefont {Wang}}, \bibinfo {author} {\bibfnamefont {Z.-P.}\ \bibnamefont
  {Yin}},\ and\ \bibinfo {author} {\bibfnamefont {S.}~\bibnamefont {Jia}},\
  }\bibfield  {title} {\bibinfo {title} {Anomalous {Hall} effect in
  ferromagnetic {LaCo$_2$As$_2$} and ferrimagnetic {NdCo$_2$As$_2$}},\ }\href
  {https://doi.org/10.1088/1674-1056/acd925} {\bibfield  {journal} {\bibinfo
  {journal} {Chin. Phys. B}\ }\textbf {\bibinfo {volume} {32}},\ \bibinfo
  {pages} {107502} (\bibinfo {year} {2023})}\BibitemShut {NoStop}%
\bibitem [{\citenamefont {Cheng}\ \emph {et~al.}(2023)\citenamefont {Cheng},
  \citenamefont {Huang}, \citenamefont {Zheng}, \citenamefont {Chen},
  \citenamefont {Cochran}, \citenamefont {Hu}, \citenamefont {Yin},
  \citenamefont {Yang}, \citenamefont {Hossain}, \citenamefont {Zhang},
  \citenamefont {Belopolski}, \citenamefont {Liu}, \citenamefont {Cheng},
  \citenamefont {Hashimoto}, \citenamefont {Lu}, \citenamefont {Xu},
  \citenamefont {Zhou}, \citenamefont {Ma}, \citenamefont {Chang},
  \citenamefont {Yao}, \citenamefont {Yin}, \citenamefont {Hasan},\ and\
  \citenamefont {Jia}}]{CCAcheng2023}%
  \BibitemOpen
  \bibfield  {author} {\bibinfo {author} {\bibfnamefont {Z.-J.}\ \bibnamefont
  {Cheng}}, \bibinfo {author} {\bibfnamefont {Y.}~\bibnamefont {Huang}},
  \bibinfo {author} {\bibfnamefont {P.}~\bibnamefont {Zheng}}, \bibinfo
  {author} {\bibfnamefont {L.}~\bibnamefont {Chen}}, \bibinfo {author}
  {\bibfnamefont {T.~A.}\ \bibnamefont {Cochran}}, \bibinfo {author}
  {\bibfnamefont {H.}~\bibnamefont {Hu}}, \bibinfo {author} {\bibfnamefont
  {J.-X.}\ \bibnamefont {Yin}}, \bibinfo {author} {\bibfnamefont {X.~P.}\
  \bibnamefont {Yang}}, \bibinfo {author} {\bibfnamefont {M.~S.}\ \bibnamefont
  {Hossain}}, \bibinfo {author} {\bibfnamefont {Q.}~\bibnamefont {Zhang}},
  \bibinfo {author} {\bibfnamefont {I.}~\bibnamefont {Belopolski}}, \bibinfo
  {author} {\bibfnamefont {R.}~\bibnamefont {Liu}}, \bibinfo {author}
  {\bibfnamefont {G.}~\bibnamefont {Cheng}}, \bibinfo {author} {\bibfnamefont
  {M.}~\bibnamefont {Hashimoto}}, \bibinfo {author} {\bibfnamefont
  {D.}~\bibnamefont {Lu}}, \bibinfo {author} {\bibfnamefont {X.}~\bibnamefont
  {Xu}}, \bibinfo {author} {\bibfnamefont {H.}~\bibnamefont {Zhou}}, \bibinfo
  {author} {\bibfnamefont {W.}~\bibnamefont {Ma}}, \bibinfo {author}
  {\bibfnamefont {G.}~\bibnamefont {Chang}}, \bibinfo {author} {\bibfnamefont
  {N.}~\bibnamefont {Yao}}, \bibinfo {author} {\bibfnamefont {Z.}~\bibnamefont
  {Yin}}, \bibinfo {author} {\bibfnamefont {M.~Z.}\ \bibnamefont {Hasan}},\
  and\ \bibinfo {author} {\bibfnamefont {S.}~\bibnamefont {Jia}},\ }\href@noop
  {} {\bibinfo {title} {Observation of {K}ondo lattice and {K}ondo-enhanced
  anomalous {H}all effect in an itinerant ferromagnet}} (\bibinfo {year}
  {2023}),\ \Eprint {https://arxiv.org/abs/2302.12113} {arXiv:2302.12113}
  \BibitemShut {NoStop}%
\bibitem [{\citenamefont {Coleman}(2007)}]{HFbook2007}%
  \BibitemOpen
  \bibfield  {author} {\bibinfo {author} {\bibfnamefont {P.}~\bibnamefont
  {Coleman}},\ }\bibinfo {title} {Heavy fermions: Electrons at the edge of
  magnetism},\ in\ \href
  {https://doi.org/https://doi.org/10.1002/9780470022184.hmm105} {\emph
  {\bibinfo {booktitle} {Handbook of Magnetism and Advanced Magnetic
  Materials}}}\ (\bibinfo  {publisher} {John Wiley \& Sons, Ltd},\ \bibinfo
  {year} {2007})\BibitemShut {NoStop}%
\bibitem [{\citenamefont {Yang}\ \emph {et~al.}(2008)\citenamefont {Yang},
  \citenamefont {Fisk}, \citenamefont {Lee}, \citenamefont {Thompson},\ and\
  \citenamefont {Pines}}]{KondoscaYang2008}%
  \BibitemOpen
  \bibfield  {author} {\bibinfo {author} {\bibfnamefont {Y.-f.}\ \bibnamefont
  {Yang}}, \bibinfo {author} {\bibfnamefont {Z.}~\bibnamefont {Fisk}}, \bibinfo
  {author} {\bibfnamefont {H.-O.}\ \bibnamefont {Lee}}, \bibinfo {author}
  {\bibfnamefont {J.~D.}\ \bibnamefont {Thompson}},\ and\ \bibinfo {author}
  {\bibfnamefont {D.}~\bibnamefont {Pines}},\ }\bibfield  {title} {\bibinfo
  {title} {Scaling the {K}ondo lattice},\ }\href
  {https://doi.org/10.1038/nature07157} {\bibfield  {journal} {\bibinfo
  {journal} {Nature}\ }\textbf {\bibinfo {volume} {454}},\ \bibinfo {pages}
  {611} (\bibinfo {year} {2008})}\BibitemShut {NoStop}%
\bibitem [{\citenamefont {Jang}\ \emph {et~al.}(2020)\citenamefont {Jang},
  \citenamefont {Denlinger}, \citenamefont {Allen}, \citenamefont {Zapf},
  \citenamefont {Maple}, \citenamefont {Kim}, \citenamefont {Jang},\ and\
  \citenamefont {Shim}}]{Jang2020evoKL}%
  \BibitemOpen
  \bibfield  {author} {\bibinfo {author} {\bibfnamefont {S.}~\bibnamefont
  {Jang}}, \bibinfo {author} {\bibfnamefont {J.~D.}\ \bibnamefont {Denlinger}},
  \bibinfo {author} {\bibfnamefont {J.~W.}\ \bibnamefont {Allen}}, \bibinfo
  {author} {\bibfnamefont {V.~S.}\ \bibnamefont {Zapf}}, \bibinfo {author}
  {\bibfnamefont {M.~B.}\ \bibnamefont {Maple}}, \bibinfo {author}
  {\bibfnamefont {J.~N.}\ \bibnamefont {Kim}}, \bibinfo {author} {\bibfnamefont
  {B.~G.}\ \bibnamefont {Jang}},\ and\ \bibinfo {author} {\bibfnamefont
  {J.~H.}\ \bibnamefont {Shim}},\ }\bibfield  {title} {\bibinfo {title}
  {Evolution of the {K}ondo lattice electronic structure above the transport
  coherence temperature},\ }\href {https://doi.org/10.1073/pnas.2001778117}
  {\bibfield  {journal} {\bibinfo  {journal} {Proc. Natl. Acad. Sci.}\ }\textbf
  {\bibinfo {volume} {117}},\ \bibinfo {pages} {23467} (\bibinfo {year}
  {2020})}\BibitemShut {NoStop}%
\bibitem [{\citenamefont {Behnia}\ \emph {et~al.}(2004)\citenamefont {Behnia},
  \citenamefont {Jaccard},\ and\ \citenamefont {Flouquet}}]{SgammaBehnia2004}%
  \BibitemOpen
  \bibfield  {author} {\bibinfo {author} {\bibfnamefont {K.}~\bibnamefont
  {Behnia}}, \bibinfo {author} {\bibfnamefont {D.}~\bibnamefont {Jaccard}},\
  and\ \bibinfo {author} {\bibfnamefont {J.}~\bibnamefont {Flouquet}},\
  }\bibfield  {title} {\bibinfo {title} {On the thermoelectricity of correlated
  electrons in the zero-temperature limit},\ }\href
  {https://doi.org/10.1088/0953-8984/16/28/037} {\bibfield  {journal} {\bibinfo
   {journal} {J. Phys. Condens. Matter}\ }\textbf {\bibinfo {volume} {16}},\
  \bibinfo {pages} {5187} (\bibinfo {year} {2004})}\BibitemShut {NoStop}%
\bibitem [{SM()}]{SM}%
  \BibitemOpen
  \bibinfo {note} {See Supplemental Material for additional methods, data, and
  analyses, which include additional Refs. \cite{b2006RMPKotliar,
  aPRB2010Haule, cBlaha2019book, dprb2018_ACo2As2, fHaule2015exact, g2007Haule,
  h2006Werner, k2012Marzari, l1996Kresse,ipizzi2020wannier90, jWU2018,
  m2010Xiao, n1982Thouless, CCAShen2014, Mangez1967Bi, Liu2018Co3Sn2S2,
  Bredl1984CeCu2Si2,Jaccard1985HF,Sparn1985Kondo,Ayache1987CePt2Si2,Sato1987CeCu6,Amato1988CeRu2Si2,Bhattacharjee1989CePt2Si2,Lacerda1989CeRu2Si2,Lohneysen1994HF,Bianchi2003CeCoIn5,Bel2004CeCoIn5}.}\BibitemShut
  {Stop}%
\bibitem [{\citenamefont {Nagaosa}\ \emph {et~al.}(2010)\citenamefont
  {Nagaosa}, \citenamefont {Sinova}, \citenamefont {Onoda}, \citenamefont
  {MacDonald},\ and\ \citenamefont {Ong}}]{RevAHE2010}%
  \BibitemOpen
  \bibfield  {author} {\bibinfo {author} {\bibfnamefont {N.}~\bibnamefont
  {Nagaosa}}, \bibinfo {author} {\bibfnamefont {J.}~\bibnamefont {Sinova}},
  \bibinfo {author} {\bibfnamefont {S.}~\bibnamefont {Onoda}}, \bibinfo
  {author} {\bibfnamefont {A.~H.}\ \bibnamefont {MacDonald}},\ and\ \bibinfo
  {author} {\bibfnamefont {N.~P.}\ \bibnamefont {Ong}},\ }\bibfield  {title}
  {\bibinfo {title} {Anomalous {H}all effect},\ }\href
  {https://doi.org/10.1103/RevModPhys.82.1539} {\bibfield  {journal} {\bibinfo
  {journal} {Rev. Mod. Phys.}\ }\textbf {\bibinfo {volume} {82}},\ \bibinfo
  {pages} {1539} (\bibinfo {year} {2010})}\BibitemShut {NoStop}%
\bibitem [{\citenamefont {Guin}\ \emph
  {et~al.}(2019{\natexlab{b}})\citenamefont {Guin}, \citenamefont {Manna},
  \citenamefont {Noky}, \citenamefont {Watzman}, \citenamefont {Fu},
  \citenamefont {Kumar}, \citenamefont {Schnelle}, \citenamefont {Shekhar},
  \citenamefont {Sun}, \citenamefont {Gooth},\ and\ \citenamefont
  {Felser}}]{Guin2019Co2MnGa}%
  \BibitemOpen
  \bibfield  {author} {\bibinfo {author} {\bibfnamefont {S.~N.}\ \bibnamefont
  {Guin}}, \bibinfo {author} {\bibfnamefont {K.}~\bibnamefont {Manna}},
  \bibinfo {author} {\bibfnamefont {J.}~\bibnamefont {Noky}}, \bibinfo {author}
  {\bibfnamefont {S.~J.}\ \bibnamefont {Watzman}}, \bibinfo {author}
  {\bibfnamefont {C.}~\bibnamefont {Fu}}, \bibinfo {author} {\bibfnamefont
  {N.}~\bibnamefont {Kumar}}, \bibinfo {author} {\bibfnamefont
  {W.}~\bibnamefont {Schnelle}}, \bibinfo {author} {\bibfnamefont
  {C.}~\bibnamefont {Shekhar}}, \bibinfo {author} {\bibfnamefont
  {Y.}~\bibnamefont {Sun}}, \bibinfo {author} {\bibfnamefont {J.}~\bibnamefont
  {Gooth}},\ and\ \bibinfo {author} {\bibfnamefont {C.}~\bibnamefont
  {Felser}},\ }\bibfield  {title} {\bibinfo {title} {Anomalous {N}ernst effect
  beyond the magnetization scaling relation in the ferromagnetic {H}eusler
  compound {C}o$_2${M}n{G}a},\ }\href
  {https://doi.org/10.1038/s41427-019-0116-z} {\bibfield  {journal} {\bibinfo
  {journal} {NPG Asia Mater.}\ }\textbf {\bibinfo {volume} {11}},\ \bibinfo
  {pages} {16} (\bibinfo {year} {2019}{\natexlab{b}})}\BibitemShut {NoStop}%
\bibitem [{\citenamefont {Xu}\ \emph {et~al.}(2020{\natexlab{b}})\citenamefont
  {Xu}, \citenamefont {Li}, \citenamefont {Ding}, \citenamefont {Chen},
  \citenamefont {Sakai}, \citenamefont {Fauqu\'e}, \citenamefont {Nakatsuji},
  \citenamefont {Zhu},\ and\ \citenamefont {Behnia}}]{Xu2020Mott}%
  \BibitemOpen
  \bibfield  {author} {\bibinfo {author} {\bibfnamefont {L.}~\bibnamefont
  {Xu}}, \bibinfo {author} {\bibfnamefont {X.}~\bibnamefont {Li}}, \bibinfo
  {author} {\bibfnamefont {L.}~\bibnamefont {Ding}}, \bibinfo {author}
  {\bibfnamefont {T.}~\bibnamefont {Chen}}, \bibinfo {author} {\bibfnamefont
  {A.}~\bibnamefont {Sakai}}, \bibinfo {author} {\bibfnamefont
  {B.}~\bibnamefont {Fauqu\'e}}, \bibinfo {author} {\bibfnamefont
  {S.}~\bibnamefont {Nakatsuji}}, \bibinfo {author} {\bibfnamefont
  {Z.}~\bibnamefont {Zhu}},\ and\ \bibinfo {author} {\bibfnamefont
  {K.}~\bibnamefont {Behnia}},\ }\bibfield  {title} {\bibinfo {title}
  {Anomalous transverse response of {C}o$_2${M}n{G}a and universality of the
  room-temperature
  ${\ensuremath{\alpha}}_{ij}^{A}/{\ensuremath{\sigma}}_{ij}^{A}$ ratio across
  topological magnets},\ }\href {https://doi.org/10.1103/PhysRevB.101.180404}
  {\bibfield  {journal} {\bibinfo  {journal} {Phys. Rev. B}\ }\textbf {\bibinfo
  {volume} {101}},\ \bibinfo {pages} {180404} (\bibinfo {year}
  {2020}{\natexlab{b}})}\BibitemShut {NoStop}%
\bibitem [{\citenamefont {Ding}\ \emph {et~al.}(2019)\citenamefont {Ding},
  \citenamefont {Koo}, \citenamefont {Xu}, \citenamefont {Li}, \citenamefont
  {Lu}, \citenamefont {Zhao}, \citenamefont {Wang}, \citenamefont {Yin},
  \citenamefont {Lei}, \citenamefont {Yan}, \citenamefont {Zhu},\ and\
  \citenamefont {Behnia}}]{Ding2019mobility}%
  \BibitemOpen
  \bibfield  {author} {\bibinfo {author} {\bibfnamefont {L.}~\bibnamefont
  {Ding}}, \bibinfo {author} {\bibfnamefont {J.}~\bibnamefont {Koo}}, \bibinfo
  {author} {\bibfnamefont {L.}~\bibnamefont {Xu}}, \bibinfo {author}
  {\bibfnamefont {X.}~\bibnamefont {Li}}, \bibinfo {author} {\bibfnamefont
  {X.}~\bibnamefont {Lu}}, \bibinfo {author} {\bibfnamefont {L.}~\bibnamefont
  {Zhao}}, \bibinfo {author} {\bibfnamefont {Q.}~\bibnamefont {Wang}}, \bibinfo
  {author} {\bibfnamefont {Q.}~\bibnamefont {Yin}}, \bibinfo {author}
  {\bibfnamefont {H.}~\bibnamefont {Lei}}, \bibinfo {author} {\bibfnamefont
  {B.}~\bibnamefont {Yan}}, \bibinfo {author} {\bibfnamefont {Z.}~\bibnamefont
  {Zhu}},\ and\ \bibinfo {author} {\bibfnamefont {K.}~\bibnamefont {Behnia}},\
  }\bibfield  {title} {\bibinfo {title} {Intrinsic anomalous {N}ernst effect
  amplified by disorder in a half-metallic semimetal},\ }\href
  {https://doi.org/10.1103/PhysRevX.9.041061} {\bibfield  {journal} {\bibinfo
  {journal} {Phys. Rev. X}\ }\textbf {\bibinfo {volume} {9}},\ \bibinfo {pages}
  {041061} (\bibinfo {year} {2019})}\BibitemShut {NoStop}%
\bibitem [{\citenamefont {Yang}\ \emph {et~al.}(2020)\citenamefont {Yang},
  \citenamefont {You}, \citenamefont {Wang}, \citenamefont {Huang},
  \citenamefont {Xi}, \citenamefont {Xu}, \citenamefont {Cao}, \citenamefont
  {Tian}, \citenamefont {Xu}, \citenamefont {Dai},\ and\ \citenamefont
  {Li}}]{Yang2020Co3Sn2S2}%
  \BibitemOpen
  \bibfield  {author} {\bibinfo {author} {\bibfnamefont {H.}~\bibnamefont
  {Yang}}, \bibinfo {author} {\bibfnamefont {W.}~\bibnamefont {You}}, \bibinfo
  {author} {\bibfnamefont {J.}~\bibnamefont {Wang}}, \bibinfo {author}
  {\bibfnamefont {J.}~\bibnamefont {Huang}}, \bibinfo {author} {\bibfnamefont
  {C.}~\bibnamefont {Xi}}, \bibinfo {author} {\bibfnamefont {X.}~\bibnamefont
  {Xu}}, \bibinfo {author} {\bibfnamefont {C.}~\bibnamefont {Cao}}, \bibinfo
  {author} {\bibfnamefont {M.}~\bibnamefont {Tian}}, \bibinfo {author}
  {\bibfnamefont {Z.-A.}\ \bibnamefont {Xu}}, \bibinfo {author} {\bibfnamefont
  {J.}~\bibnamefont {Dai}},\ and\ \bibinfo {author} {\bibfnamefont
  {Y.}~\bibnamefont {Li}},\ }\bibfield  {title} {\bibinfo {title} {Giant
  anomalous {N}ernst effect in the magnetic weyl semimetal
  {C}o$_3${S}n$_2${S}$_2$},\ }\href
  {https://doi.org/10.1103/PhysRevMaterials.4.024202} {\bibfield  {journal}
  {\bibinfo  {journal} {Phys. Rev. Mater.}\ }\textbf {\bibinfo {volume} {4}},\
  \bibinfo {pages} {024202} (\bibinfo {year} {2020})}\BibitemShut {NoStop}%
\bibitem [{\citenamefont {Guo}\ \emph {et~al.}(2023)\citenamefont {Guo},
  \citenamefont {Li}, \citenamefont {Zhu},\ and\ \citenamefont
  {Behnia}}]{Guo2023Onsager}%
  \BibitemOpen
  \bibfield  {author} {\bibinfo {author} {\bibfnamefont {X.}~\bibnamefont
  {Guo}}, \bibinfo {author} {\bibfnamefont {X.}~\bibnamefont {Li}}, \bibinfo
  {author} {\bibfnamefont {Z.}~\bibnamefont {Zhu}},\ and\ \bibinfo {author}
  {\bibfnamefont {K.}~\bibnamefont {Behnia}},\ }\bibfield  {title} {\bibinfo
  {title} {Onsager reciprocal relation between anomalous transverse
  coefficients of an anisotropic antiferromagnet},\ }\href
  {https://doi.org/10.1103/PhysRevLett.131.246302} {\bibfield  {journal}
  {\bibinfo  {journal} {Phys. Rev. Lett.}\ }\textbf {\bibinfo {volume} {131}},\
  \bibinfo {pages} {246302} (\bibinfo {year} {2023})}\BibitemShut {NoStop}%
\bibitem [{\citenamefont {Tian}\ \emph {et~al.}(2009)\citenamefont {Tian},
  \citenamefont {Ye},\ and\ \citenamefont {Jin}}]{tian2009proper}%
  \BibitemOpen
  \bibfield  {author} {\bibinfo {author} {\bibfnamefont {Y.}~\bibnamefont
  {Tian}}, \bibinfo {author} {\bibfnamefont {L.}~\bibnamefont {Ye}},\ and\
  \bibinfo {author} {\bibfnamefont {X.}~\bibnamefont {Jin}},\ }\bibfield
  {title} {\bibinfo {title} {Proper scaling of the anomalous {H}all effect},\
  }\href {https://doi.org/10.1103/PhysRevLett.103.087206} {\bibfield  {journal}
  {\bibinfo  {journal} {Phys. Rev. Lett.}\ }\textbf {\bibinfo {volume} {103}},\
  \bibinfo {pages} {087206} (\bibinfo {year} {2009})}\BibitemShut {NoStop}%
\bibitem [{\citenamefont {Onoda}\ \emph {et~al.}(2008)\citenamefont {Onoda},
  \citenamefont {Sugimoto},\ and\ \citenamefont {Nagaosa}}]{Onoda2008quantum}%
  \BibitemOpen
  \bibfield  {author} {\bibinfo {author} {\bibfnamefont {S.}~\bibnamefont
  {Onoda}}, \bibinfo {author} {\bibfnamefont {N.}~\bibnamefont {Sugimoto}},\
  and\ \bibinfo {author} {\bibfnamefont {N.}~\bibnamefont {Nagaosa}},\
  }\bibfield  {title} {\bibinfo {title} {Quantum transport theory of anomalous
  electric, thermoelectric, and thermal {H}all effects in ferromagnets},\
  }\href {https://doi.org/10.1103/PhysRevB.77.165103} {\bibfield  {journal}
  {\bibinfo  {journal} {Phys. Rev. B}\ }\textbf {\bibinfo {volume} {77}},\
  \bibinfo {pages} {165103} (\bibinfo {year} {2008})}\BibitemShut {NoStop}%
\bibitem [{\citenamefont {Nair}\ \emph {et~al.}(2012)\citenamefont {Nair},
  \citenamefont {Wirth}, \citenamefont {Friedemann}, \citenamefont {Steglich},
  \citenamefont {Si},\ and\ \citenamefont {Schofield}}]{Nair2012HEinHF}%
  \BibitemOpen
  \bibfield  {author} {\bibinfo {author} {\bibfnamefont {S.}~\bibnamefont
  {Nair}}, \bibinfo {author} {\bibfnamefont {S.}~\bibnamefont {Wirth}},
  \bibinfo {author} {\bibfnamefont {S.}~\bibnamefont {Friedemann}}, \bibinfo
  {author} {\bibfnamefont {F.}~\bibnamefont {Steglich}}, \bibinfo {author}
  {\bibfnamefont {Q.}~\bibnamefont {Si}},\ and\ \bibinfo {author}
  {\bibfnamefont {A.~J.}\ \bibnamefont {Schofield}},\ }\bibfield  {title}
  {\bibinfo {title} {Hall effect in heavy fermion metals},\ }\href
  {https://doi.org/10.1080/00018732.2012.730223} {\bibfield  {journal}
  {\bibinfo  {journal} {Adv. Phys.}\ }\textbf {\bibinfo {volume} {61}},\
  \bibinfo {pages} {583} (\bibinfo {year} {2012})}\BibitemShut {NoStop}%
\bibitem [{\citenamefont {feng Yang}(2016)}]{Yang2016twofluid}%
  \BibitemOpen
  \bibfield  {author} {\bibinfo {author} {\bibfnamefont {Y.}~\bibnamefont {feng
  Yang}},\ }\bibfield  {title} {\bibinfo {title} {Two-fluid model for heavy
  electron physics},\ }\href {https://doi.org/10.1088/0034-4885/79/7/074501}
  {\bibfield  {journal} {\bibinfo  {journal} {Rep. Prog. Phys.}\ }\textbf
  {\bibinfo {volume} {79}},\ \bibinfo {pages} {074501} (\bibinfo {year}
  {2016})}\BibitemShut {NoStop}%
\bibitem [{\citenamefont {Noky}\ \emph
  {et~al.}(2018{\natexlab{b}})\citenamefont {Noky}, \citenamefont {Gooth},
  \citenamefont {Felser},\ and\ \citenamefont {Sun}}]{Sun2018away}%
  \BibitemOpen
  \bibfield  {author} {\bibinfo {author} {\bibfnamefont {J.}~\bibnamefont
  {Noky}}, \bibinfo {author} {\bibfnamefont {J.}~\bibnamefont {Gooth}},
  \bibinfo {author} {\bibfnamefont {C.}~\bibnamefont {Felser}},\ and\ \bibinfo
  {author} {\bibfnamefont {Y.}~\bibnamefont {Sun}},\ }\bibfield  {title}
  {\bibinfo {title} {Characterization of topological band structures away from
  the {F}ermi level by the anomalous {N}ernst effect},\ }\href
  {https://doi.org/10.1103/PhysRevB.98.241106} {\bibfield  {journal} {\bibinfo
  {journal} {Phys. Rev. B}\ }\textbf {\bibinfo {volume} {98}},\ \bibinfo
  {pages} {241106} (\bibinfo {year} {2018}{\natexlab{b}})}\BibitemShut
  {NoStop}%
\bibitem [{\citenamefont {Behnia}(2015)}]{2015FundamentalsBehnia}%
  \BibitemOpen
  \bibfield  {author} {\bibinfo {author} {\bibfnamefont {K.}~\bibnamefont
  {Behnia}},\ }\href
  {https://doi.org/10.1093/acprof:oso/9780199697663.001.0001} {\emph {\bibinfo
  {title} {{Fundamentals of Thermoelectricity}}}}\ (\bibinfo  {publisher}
  {Oxford University Press},\ \bibinfo {year} {2015})\BibitemShut {NoStop}%
\bibitem [{\citenamefont {Behnia}(2022)}]{Behnia2022TEP}%
  \BibitemOpen
  \bibfield  {author} {\bibinfo {author} {\bibfnamefont {K.}~\bibnamefont
  {Behnia}},\ }\bibfield  {title} {\bibinfo {title} {What is measured when
  measuring a thermoelectric coefficient?},\ }\href
  {https://doi.org/10.5802/crphys.100} {\bibfield  {journal} {\bibinfo
  {journal} {C. R. Phys.}\ }\textbf {\bibinfo {volume} {23}},\ \bibinfo {pages}
  {25} (\bibinfo {year} {2022})}\BibitemShut {NoStop}%
\bibitem [{\citenamefont {Qiang}\ \emph {et~al.}(2023)\citenamefont {Qiang},
  \citenamefont {Du}, \citenamefont {Lu},\ and\ \citenamefont
  {Xie}}]{Lu2023Mott}%
  \BibitemOpen
  \bibfield  {author} {\bibinfo {author} {\bibfnamefont {X.-B.}\ \bibnamefont
  {Qiang}}, \bibinfo {author} {\bibfnamefont {Z.~Z.}\ \bibnamefont {Du}},
  \bibinfo {author} {\bibfnamefont {H.-Z.}\ \bibnamefont {Lu}},\ and\ \bibinfo
  {author} {\bibfnamefont {X.~C.}\ \bibnamefont {Xie}},\ }\bibfield  {title}
  {\bibinfo {title} {Topological and disorder corrections to the transverse
  {W}iedemann-{F}ranz law and {M}ott relation in kagome magnets and {D}irac
  materials},\ }\href {https://doi.org/10.1103/PhysRevB.107.L161302} {\bibfield
   {journal} {\bibinfo  {journal} {Phys. Rev. B}\ }\textbf {\bibinfo {volume}
  {107}},\ \bibinfo {pages} {L161302} (\bibinfo {year} {2023})}\BibitemShut
  {NoStop}%
\bibitem [{\citenamefont {Miyasato}\ \emph {et~al.}(2007)\citenamefont
  {Miyasato}, \citenamefont {Abe}, \citenamefont {Fujii}, \citenamefont
  {Asamitsu}, \citenamefont {Onoda}, \citenamefont {Onose}, \citenamefont
  {Nagaosa},\ and\ \citenamefont {Tokura}}]{crossoverbehaviorTukura2007}%
  \BibitemOpen
  \bibfield  {author} {\bibinfo {author} {\bibfnamefont {T.}~\bibnamefont
  {Miyasato}}, \bibinfo {author} {\bibfnamefont {N.}~\bibnamefont {Abe}},
  \bibinfo {author} {\bibfnamefont {T.}~\bibnamefont {Fujii}}, \bibinfo
  {author} {\bibfnamefont {A.}~\bibnamefont {Asamitsu}}, \bibinfo {author}
  {\bibfnamefont {S.}~\bibnamefont {Onoda}}, \bibinfo {author} {\bibfnamefont
  {Y.}~\bibnamefont {Onose}}, \bibinfo {author} {\bibfnamefont
  {N.}~\bibnamefont {Nagaosa}},\ and\ \bibinfo {author} {\bibfnamefont
  {Y.}~\bibnamefont {Tokura}},\ }\bibfield  {title} {\bibinfo {title}
  {Crossover behavior of the anomalous {H}all effect and anomalous {N}ernst
  effect in itinerant ferromagnets},\ }\href
  {https://doi.org/10.1103/PhysRevLett.99.086602} {\bibfield  {journal}
  {\bibinfo  {journal} {Phys. Rev. Lett.}\ }\textbf {\bibinfo {volume} {99}},\
  \bibinfo {pages} {086602} (\bibinfo {year} {2007})}\BibitemShut {NoStop}%
\bibitem [{\citenamefont {Pu}\ \emph {et~al.}(2008)\citenamefont {Pu},
  \citenamefont {Chiba}, \citenamefont {Matsukura}, \citenamefont {Ohno},\ and\
  \citenamefont {Shi}}]{Pu2008GaMnAs}%
  \BibitemOpen
  \bibfield  {author} {\bibinfo {author} {\bibfnamefont {Y.}~\bibnamefont
  {Pu}}, \bibinfo {author} {\bibfnamefont {D.}~\bibnamefont {Chiba}}, \bibinfo
  {author} {\bibfnamefont {F.}~\bibnamefont {Matsukura}}, \bibinfo {author}
  {\bibfnamefont {H.}~\bibnamefont {Ohno}},\ and\ \bibinfo {author}
  {\bibfnamefont {J.}~\bibnamefont {Shi}},\ }\bibfield  {title} {\bibinfo
  {title} {Mott relation for anomalous {H}all and {N}ernst effects in
  {G}a$_\mathrm{1-x}${M}n$_\mathrm{x}${A}s ferromagnetic semiconductors},\
  }\href {https://doi.org/10.1103/PhysRevLett.101.117208} {\bibfield  {journal}
  {\bibinfo  {journal} {Phys. Rev. Lett.}\ }\textbf {\bibinfo {volume} {101}},\
  \bibinfo {pages} {117208} (\bibinfo {year} {2008})}\BibitemShut {NoStop}%
\bibitem [{\citenamefont {Xu}\ \emph {et~al.}(2022)\citenamefont {Xu},
  \citenamefont {Yin}, \citenamefont {Ma}, \citenamefont {Tien}, \citenamefont
  {Qiang}, \citenamefont {Reddy}, \citenamefont {Zhou}, \citenamefont {Shen},
  \citenamefont {Lu}, \citenamefont {Chang}, \citenamefont {Qu},\ and\
  \citenamefont {Jia}}]{Xu2022Tb166}%
  \BibitemOpen
  \bibfield  {author} {\bibinfo {author} {\bibfnamefont {X.}~\bibnamefont
  {Xu}}, \bibinfo {author} {\bibfnamefont {J.-X.}\ \bibnamefont {Yin}},
  \bibinfo {author} {\bibfnamefont {W.}~\bibnamefont {Ma}}, \bibinfo {author}
  {\bibfnamefont {H.-J.}\ \bibnamefont {Tien}}, \bibinfo {author}
  {\bibfnamefont {X.-B.}\ \bibnamefont {Qiang}}, \bibinfo {author}
  {\bibfnamefont {P.~V.~S.}\ \bibnamefont {Reddy}}, \bibinfo {author}
  {\bibfnamefont {H.}~\bibnamefont {Zhou}}, \bibinfo {author} {\bibfnamefont
  {J.}~\bibnamefont {Shen}}, \bibinfo {author} {\bibfnamefont {H.-Z.}\
  \bibnamefont {Lu}}, \bibinfo {author} {\bibfnamefont {T.-R.}\ \bibnamefont
  {Chang}}, \bibinfo {author} {\bibfnamefont {Z.}~\bibnamefont {Qu}},\ and\
  \bibinfo {author} {\bibfnamefont {S.}~\bibnamefont {Jia}},\ }\bibfield
  {title} {\bibinfo {title} {Topological charge-entropy scaling in kagome chern
  magnet {TbMn$_6$Sn$_6$}},\ }\href
  {https://doi.org/10.1038/s41467-022-28796-6} {\bibfield  {journal} {\bibinfo
  {journal} {Nat. Commun.}\ }\textbf {\bibinfo {volume} {13}},\ \bibinfo
  {pages} {1197} (\bibinfo {year} {2022})}\BibitemShut {NoStop}%
\bibitem [{\citenamefont {Lai}\ \emph {et~al.}(2018)\citenamefont {Lai},
  \citenamefont {Grefe}, \citenamefont {Paschen},\ and\ \citenamefont
  {Si}}]{lai_weylkondo_2018}%
  \BibitemOpen
  \bibfield  {author} {\bibinfo {author} {\bibfnamefont {H.-H.}\ \bibnamefont
  {Lai}}, \bibinfo {author} {\bibfnamefont {S.~E.}\ \bibnamefont {Grefe}},
  \bibinfo {author} {\bibfnamefont {S.}~\bibnamefont {Paschen}},\ and\ \bibinfo
  {author} {\bibfnamefont {Q.}~\bibnamefont {Si}},\ }\bibfield  {title}
  {\bibinfo {title} {Weyl–{Kondo} semimetal in heavy-fermion systems},\
  }\href {https://www.pnas.org/doi/abs/10.1073/pnas.1715851115} {\bibfield
  {journal} {\bibinfo  {journal} {Proc. Natl. Acad. Sci.}\ }\textbf {\bibinfo
  {volume} {115}},\ \bibinfo {pages} {93} (\bibinfo {year} {2018})}\BibitemShut
  {NoStop}%
\bibitem [{\citenamefont {Grefe}\ \emph {et~al.}(2020)\citenamefont {Grefe},
  \citenamefont {Lai}, \citenamefont {Paschen},\ and\ \citenamefont
  {Si}}]{WeylKS2019}%
  \BibitemOpen
  \bibfield  {author} {\bibinfo {author} {\bibfnamefont {S.~E.}\ \bibnamefont
  {Grefe}}, \bibinfo {author} {\bibfnamefont {H.-H.}\ \bibnamefont {Lai}},
  \bibinfo {author} {\bibfnamefont {S.}~\bibnamefont {Paschen}},\ and\ \bibinfo
  {author} {\bibfnamefont {Q.}~\bibnamefont {Si}},\ }\bibinfo {title}
  {Weyl–{K}ondo semimetal: Towards control of {W}eyl nodes},\ in\ \href
  {https://doi.org/10.7566/JPSCP.30.011013} {\emph {\bibinfo {booktitle}
  {Proceedings of the International Conference on Strongly Correlated Electron
  Systems (SCES2019)}}}\ (\bibinfo {year} {2020})\ p.\ \bibinfo {pages}
  {011013}\BibitemShut {NoStop}%
\bibitem [{\citenamefont {Paschen}\ and\ \citenamefont
  {Si}(2021)}]{Paschen2021topoKondo}%
  \BibitemOpen
  \bibfield  {author} {\bibinfo {author} {\bibfnamefont {S.}~\bibnamefont
  {Paschen}}\ and\ \bibinfo {author} {\bibfnamefont {Q.}~\bibnamefont {Si}},\
  }\bibfield  {title} {\bibinfo {title} {Quantum phases driven by strong
  correlations},\ }\href {https://doi.org/10.1038/s42254-020-00262-6}
  {\bibfield  {journal} {\bibinfo  {journal} {Nat. Rev. Phys.}\ }\textbf
  {\bibinfo {volume} {3}},\ \bibinfo {pages} {9} (\bibinfo {year}
  {2021})}\BibitemShut {NoStop}%
\bibitem [{\citenamefont {Chen}\ \emph
  {et~al.}(2022{\natexlab{b}})\citenamefont {Chen}, \citenamefont {Setty},
  \citenamefont {Hu}, \citenamefont {Vergniory}, \citenamefont {Grefe},
  \citenamefont {Fischer}, \citenamefont {Yan}, \citenamefont {Eguchi},
  \citenamefont {Prokofiev}, \citenamefont {Paschen}, \citenamefont {Cano},\
  and\ \citenamefont {Si}}]{Chen2022topocorre}%
  \BibitemOpen
  \bibfield  {author} {\bibinfo {author} {\bibfnamefont {L.}~\bibnamefont
  {Chen}}, \bibinfo {author} {\bibfnamefont {C.}~\bibnamefont {Setty}},
  \bibinfo {author} {\bibfnamefont {H.}~\bibnamefont {Hu}}, \bibinfo {author}
  {\bibfnamefont {M.~G.}\ \bibnamefont {Vergniory}}, \bibinfo {author}
  {\bibfnamefont {S.~E.}\ \bibnamefont {Grefe}}, \bibinfo {author}
  {\bibfnamefont {L.}~\bibnamefont {Fischer}}, \bibinfo {author} {\bibfnamefont
  {X.}~\bibnamefont {Yan}}, \bibinfo {author} {\bibfnamefont {G.}~\bibnamefont
  {Eguchi}}, \bibinfo {author} {\bibfnamefont {A.}~\bibnamefont {Prokofiev}},
  \bibinfo {author} {\bibfnamefont {S.}~\bibnamefont {Paschen}}, \bibinfo
  {author} {\bibfnamefont {J.}~\bibnamefont {Cano}},\ and\ \bibinfo {author}
  {\bibfnamefont {Q.}~\bibnamefont {Si}},\ }\bibfield  {title} {\bibinfo
  {title} {Topological semimetal driven by strong correlations and crystalline
  symmetry},\ }\href {https://doi.org/10.1038/s41567-022-01743-4} {\bibfield
  {journal} {\bibinfo  {journal} {Nat. Phys.}\ }\textbf {\bibinfo {volume}
  {18}},\ \bibinfo {pages} {1341} (\bibinfo {year}
  {2022}{\natexlab{b}})}\BibitemShut {NoStop}%
\bibitem [{\citenamefont {Checkelsky}\ \emph {et~al.}(2024)\citenamefont
  {Checkelsky}, \citenamefont {Bernevig}, \citenamefont {Coleman},
  \citenamefont {Si},\ and\ \citenamefont {Paschen}}]{Checkelsky2024flatband}%
  \BibitemOpen
  \bibfield  {author} {\bibinfo {author} {\bibfnamefont {J.~G.}\ \bibnamefont
  {Checkelsky}}, \bibinfo {author} {\bibfnamefont {B.~A.}\ \bibnamefont
  {Bernevig}}, \bibinfo {author} {\bibfnamefont {P.}~\bibnamefont {Coleman}},
  \bibinfo {author} {\bibfnamefont {Q.}~\bibnamefont {Si}},\ and\ \bibinfo
  {author} {\bibfnamefont {S.}~\bibnamefont {Paschen}},\ }\bibfield  {title}
  {\bibinfo {title} {Flat bands, strange metals and the {K}ondo effect},\
  }\href {https://doi.org/10.1038/s41578-023-00644-z} {\bibfield  {journal}
  {\bibinfo  {journal} {Nat. Rev. Mater.}\ }\textbf {\bibinfo {volume} {9}},\
  \bibinfo {pages} {509} (\bibinfo {year} {2024})}\BibitemShut {NoStop}%
\bibitem [{\citenamefont {Dzsaber}\ \emph {et~al.}(2021)\citenamefont
  {Dzsaber}, \citenamefont {Yan}, \citenamefont {Taupin}, \citenamefont
  {Eguchi}, \citenamefont {Prokofiev}, \citenamefont {Shiroka}, \citenamefont
  {Blaha}, \citenamefont {Rubel}, \citenamefont {Grefe}, \citenamefont {Lai},
  \citenamefont {Si},\ and\ \citenamefont {Paschen}}]{dzsaber_giant_2021}%
  \BibitemOpen
  \bibfield  {author} {\bibinfo {author} {\bibfnamefont {S.}~\bibnamefont
  {Dzsaber}}, \bibinfo {author} {\bibfnamefont {X.}~\bibnamefont {Yan}},
  \bibinfo {author} {\bibfnamefont {M.}~\bibnamefont {Taupin}}, \bibinfo
  {author} {\bibfnamefont {G.}~\bibnamefont {Eguchi}}, \bibinfo {author}
  {\bibfnamefont {A.}~\bibnamefont {Prokofiev}}, \bibinfo {author}
  {\bibfnamefont {T.}~\bibnamefont {Shiroka}}, \bibinfo {author} {\bibfnamefont
  {P.}~\bibnamefont {Blaha}}, \bibinfo {author} {\bibfnamefont
  {O.}~\bibnamefont {Rubel}}, \bibinfo {author} {\bibfnamefont {S.~E.}\
  \bibnamefont {Grefe}}, \bibinfo {author} {\bibfnamefont {H.-H.}\ \bibnamefont
  {Lai}}, \bibinfo {author} {\bibfnamefont {Q.}~\bibnamefont {Si}},\ and\
  \bibinfo {author} {\bibfnamefont {S.}~\bibnamefont {Paschen}},\ }\bibfield
  {title} {\bibinfo {title} {Giant spontaneous {H}all effect in a nonmagnetic
  {W}eyl–{K}ondo semimetal},\ }\href
  {https://doi.org/10.1073/pnas.2013386118} {\bibfield  {journal} {\bibinfo
  {journal} {Proc. Natl. Acad. Sci.}\ }\textbf {\bibinfo {volume} {118}},\
  \bibinfo {pages} {e2013386118} (\bibinfo {year} {2021})}\BibitemShut
  {NoStop}%
\bibitem [{\citenamefont {Kotliar}\ \emph {et~al.}(2006)\citenamefont
  {Kotliar}, \citenamefont {Savrasov}, \citenamefont {Haule}, \citenamefont
  {Oudovenko}, \citenamefont {Parcollet},\ and\ \citenamefont
  {Marianetti}}]{b2006RMPKotliar}%
  \BibitemOpen
  \bibfield  {author} {\bibinfo {author} {\bibfnamefont {G.}~\bibnamefont
  {Kotliar}}, \bibinfo {author} {\bibfnamefont {S.~Y.}\ \bibnamefont
  {Savrasov}}, \bibinfo {author} {\bibfnamefont {K.}~\bibnamefont {Haule}},
  \bibinfo {author} {\bibfnamefont {V.~S.}\ \bibnamefont {Oudovenko}}, \bibinfo
  {author} {\bibfnamefont {O.}~\bibnamefont {Parcollet}},\ and\ \bibinfo
  {author} {\bibfnamefont {C.~A.}\ \bibnamefont {Marianetti}},\ }\bibfield
  {title} {\bibinfo {title} {Electronic structure calculations with dynamical
  mean-field theory},\ }\href {https://doi.org/10.1103/RevModPhys.78.865}
  {\bibfield  {journal} {\bibinfo  {journal} {Rev. Mod. Phys.}\ }\textbf
  {\bibinfo {volume} {78}},\ \bibinfo {pages} {865} (\bibinfo {year}
  {2006})}\BibitemShut {NoStop}%
\bibitem [{\citenamefont {Haule}\ \emph {et~al.}(2010)\citenamefont {Haule},
  \citenamefont {Yee},\ and\ \citenamefont {Kim}}]{aPRB2010Haule}%
  \BibitemOpen
  \bibfield  {author} {\bibinfo {author} {\bibfnamefont {K.}~\bibnamefont
  {Haule}}, \bibinfo {author} {\bibfnamefont {C.-H.}\ \bibnamefont {Yee}},\
  and\ \bibinfo {author} {\bibfnamefont {K.}~\bibnamefont {Kim}},\ }\bibfield
  {title} {\bibinfo {title} {Dynamical mean-field theory within the
  full-potential methods: Electronic structure of {CeIrIn$_5$, CeCoIn$_5$, and
  CeRhIn$_5$}},\ }\href {https://doi.org/10.1103/PhysRevB.81.195107} {\bibfield
   {journal} {\bibinfo  {journal} {Phys. Rev. B}\ }\textbf {\bibinfo {volume}
  {81}},\ \bibinfo {pages} {195107} (\bibinfo {year} {2010})}\BibitemShut
  {NoStop}%
\bibitem [{\citenamefont {Blaha}\ \emph {et~al.}(2019)\citenamefont {Blaha},
  \citenamefont {Schwarz}, \citenamefont {Madsen}, \citenamefont {Kvasnicka},
  \citenamefont {Luitz}, \citenamefont {Laskowsk}, \citenamefont {Tran},\ and\
  \citenamefont {Marks}}]{cBlaha2019book}%
  \BibitemOpen
  \bibfield  {author} {\bibinfo {author} {\bibfnamefont {P.}~\bibnamefont
  {Blaha}}, \bibinfo {author} {\bibfnamefont {K.}~\bibnamefont {Schwarz}},
  \bibinfo {author} {\bibfnamefont {G.~K.~H.}\ \bibnamefont {Madsen}}, \bibinfo
  {author} {\bibfnamefont {D.}~\bibnamefont {Kvasnicka}}, \bibinfo {author}
  {\bibfnamefont {J.}~\bibnamefont {Luitz}}, \bibinfo {author} {\bibfnamefont
  {R.}~\bibnamefont {Laskowsk}}, \bibinfo {author} {\bibfnamefont
  {F.}~\bibnamefont {Tran}},\ and\ \bibinfo {author} {\bibfnamefont
  {L.}~\bibnamefont {Marks}},\ }\href
  {http://susi.theochem.tuwien.ac.at/reg_user/textbooks/usersguide.pdf} {\emph
  {\bibinfo {title} {WIEN2k: An Augmented Plane Wave Plus Local Orbitals
  Program for Calculating Crystal Properties}}},\ \bibinfo {edition} {19th}\
  ed.\ (\bibinfo  {publisher} {Technische Universit{\"a}t Wien},\ \bibinfo
  {year} {2019})\BibitemShut {NoStop}%
\bibitem [{\citenamefont {Mao}\ and\ \citenamefont
  {Yin}(2018)}]{dprb2018_ACo2As2}%
  \BibitemOpen
  \bibfield  {author} {\bibinfo {author} {\bibfnamefont {H.}~\bibnamefont
  {Mao}}\ and\ \bibinfo {author} {\bibfnamefont {Z.}~\bibnamefont {Yin}},\
  }\bibfield  {title} {\bibinfo {title} {{Electronic structure and spin
  dynamics of ACo\textsubscript{2}As\textsubscript{2}(A = Ba, Sr, Ca)}},\
  }\href {https://doi.org/10.1103/PhysRevB.98.115128} {\bibfield  {journal}
  {\bibinfo  {journal} {Phys. Rev. B}\ }\textbf {\bibinfo {volume} {98}},\
  \bibinfo {pages} {115128} (\bibinfo {year} {2018})}\BibitemShut {NoStop}%
\bibitem [{\citenamefont {Haule}(2015)}]{fHaule2015exact}%
  \BibitemOpen
  \bibfield  {author} {\bibinfo {author} {\bibfnamefont {K.}~\bibnamefont
  {Haule}},\ }\bibfield  {title} {\bibinfo {title} {Exact double counting in
  combining the dynamical mean field theory and the density functional
  theory},\ }\href {https://doi.org/10.1103/PhysRevLett.115.196403} {\bibfield
  {journal} {\bibinfo  {journal} {Phys. Rev. Lett.}\ }\textbf {\bibinfo
  {volume} {115}},\ \bibinfo {pages} {196403} (\bibinfo {year}
  {2015})}\BibitemShut {NoStop}%
\bibitem [{\citenamefont {Haule}(2007)}]{g2007Haule}%
  \BibitemOpen
  \bibfield  {author} {\bibinfo {author} {\bibfnamefont {K.}~\bibnamefont
  {Haule}},\ }\bibfield  {title} {\bibinfo {title} {Quantum monte carlo
  impurity solver for cluster dynamical mean-field theory and electronic
  structure calculations with adjustable cluster base},\ }\href
  {https://doi.org/10.1103/PhysRevB.75.155113} {\bibfield  {journal} {\bibinfo
  {journal} {Phys. Rev. B}\ }\textbf {\bibinfo {volume} {75}},\ \bibinfo
  {pages} {155113} (\bibinfo {year} {2007})}\BibitemShut {NoStop}%
\bibitem [{\citenamefont {Werner}\ \emph {et~al.}(2006)\citenamefont {Werner},
  \citenamefont {Comanac}, \citenamefont {de' Medici}, \citenamefont {Troyer},\
  and\ \citenamefont {Millis}}]{h2006Werner}%
  \BibitemOpen
  \bibfield  {author} {\bibinfo {author} {\bibfnamefont {P.}~\bibnamefont
  {Werner}}, \bibinfo {author} {\bibfnamefont {A.}~\bibnamefont {Comanac}},
  \bibinfo {author} {\bibfnamefont {L.}~\bibnamefont {de' Medici}}, \bibinfo
  {author} {\bibfnamefont {M.}~\bibnamefont {Troyer}},\ and\ \bibinfo {author}
  {\bibfnamefont {A.~J.}\ \bibnamefont {Millis}},\ }\bibfield  {title}
  {\bibinfo {title} {Continuous-time solver for quantum impurity models},\
  }\href {https://doi.org/10.1103/PhysRevLett.97.076405} {\bibfield  {journal}
  {\bibinfo  {journal} {Phys. Rev. Lett.}\ }\textbf {\bibinfo {volume} {97}},\
  \bibinfo {pages} {076405} (\bibinfo {year} {2006})}\BibitemShut {NoStop}%
\bibitem [{\citenamefont {Marzari}\ \emph {et~al.}(2012)\citenamefont
  {Marzari}, \citenamefont {Mostofi}, \citenamefont {Yates}, \citenamefont
  {Souza},\ and\ \citenamefont {Vanderbilt}}]{k2012Marzari}%
  \BibitemOpen
  \bibfield  {author} {\bibinfo {author} {\bibfnamefont {N.}~\bibnamefont
  {Marzari}}, \bibinfo {author} {\bibfnamefont {A.~A.}\ \bibnamefont
  {Mostofi}}, \bibinfo {author} {\bibfnamefont {J.~R.}\ \bibnamefont {Yates}},
  \bibinfo {author} {\bibfnamefont {I.}~\bibnamefont {Souza}},\ and\ \bibinfo
  {author} {\bibfnamefont {D.}~\bibnamefont {Vanderbilt}},\ }\bibfield  {title}
  {\bibinfo {title} {Maximally localized wannier functions: Theory and
  applications},\ }\href {https://doi.org/10.1103/RevModPhys.84.1419}
  {\bibfield  {journal} {\bibinfo  {journal} {Rev. Mod. Phys.}\ }\textbf
  {\bibinfo {volume} {84}},\ \bibinfo {pages} {1419} (\bibinfo {year}
  {2012})}\BibitemShut {NoStop}%
\bibitem [{\citenamefont {Kresse}\ and\ \citenamefont
  {Furthmüller}(1996)}]{l1996Kresse}%
  \BibitemOpen
  \bibfield  {author} {\bibinfo {author} {\bibfnamefont {G.}~\bibnamefont
  {Kresse}}\ and\ \bibinfo {author} {\bibfnamefont {J.}~\bibnamefont
  {Furthmüller}},\ }\bibfield  {title} {\bibinfo {title} {Efficiency of
  ab-initio total energy calculations for metals and semiconductors using a
  plane-wave basis set},\ }\href
  {https://doi.org/https://doi.org/10.1016/0927-0256(96)00008-0} {\bibfield
  {journal} {\bibinfo  {journal} {Comput. Mater. Sci.}\ }\textbf {\bibinfo
  {volume} {6}},\ \bibinfo {pages} {15} (\bibinfo {year} {1996})}\BibitemShut
  {NoStop}%
\bibitem [{\citenamefont {Pizzi}\ \emph {et~al.}(2020)\citenamefont {Pizzi},
  \citenamefont {Vitale}, \citenamefont {Arita}, \citenamefont {Bl{\"u}gel},
  \citenamefont {Freimuth}, \citenamefont {G{\'e}ranton}, \citenamefont
  {Gibertini}, \citenamefont {Gresch}, \citenamefont {Johnson}, \citenamefont
  {Koretsune} \emph {et~al.}}]{ipizzi2020wannier90}%
  \BibitemOpen
  \bibfield  {author} {\bibinfo {author} {\bibfnamefont {G.}~\bibnamefont
  {Pizzi}}, \bibinfo {author} {\bibfnamefont {V.}~\bibnamefont {Vitale}},
  \bibinfo {author} {\bibfnamefont {R.}~\bibnamefont {Arita}}, \bibinfo
  {author} {\bibfnamefont {S.}~\bibnamefont {Bl{\"u}gel}}, \bibinfo {author}
  {\bibfnamefont {F.}~\bibnamefont {Freimuth}}, \bibinfo {author}
  {\bibfnamefont {G.}~\bibnamefont {G{\'e}ranton}}, \bibinfo {author}
  {\bibfnamefont {M.}~\bibnamefont {Gibertini}}, \bibinfo {author}
  {\bibfnamefont {D.}~\bibnamefont {Gresch}}, \bibinfo {author} {\bibfnamefont
  {C.}~\bibnamefont {Johnson}}, \bibinfo {author} {\bibfnamefont
  {T.}~\bibnamefont {Koretsune}}, \emph {et~al.},\ }\bibfield  {title}
  {\bibinfo {title} {Wannier90 as a community code: new features and
  applications},\ }\href@noop {} {\bibfield  {journal} {\bibinfo  {journal}
  {Journal of Physics: Condensed Matter}\ }\textbf {\bibinfo {volume} {32}},\
  \bibinfo {pages} {165902} (\bibinfo {year} {2020})}\BibitemShut {NoStop}%
\bibitem [{\citenamefont {Wu}\ \emph {et~al.}(2018)\citenamefont {Wu},
  \citenamefont {Zhang}, \citenamefont {Song}, \citenamefont {Troyer},\ and\
  \citenamefont {Soluyanov}}]{jWU2018}%
  \BibitemOpen
  \bibfield  {author} {\bibinfo {author} {\bibfnamefont {Q.}~\bibnamefont
  {Wu}}, \bibinfo {author} {\bibfnamefont {S.}~\bibnamefont {Zhang}}, \bibinfo
  {author} {\bibfnamefont {H.-F.}\ \bibnamefont {Song}}, \bibinfo {author}
  {\bibfnamefont {M.}~\bibnamefont {Troyer}},\ and\ \bibinfo {author}
  {\bibfnamefont {A.~A.}\ \bibnamefont {Soluyanov}},\ }\bibfield  {title}
  {\bibinfo {title} {Wanniertools: An open-source software package for novel
  topological materials},\ }\href
  {https://doi.org/https://doi.org/10.1016/j.cpc.2017.09.033} {\bibfield
  {journal} {\bibinfo  {journal} {Comput. Phys. Commun.}\ }\textbf {\bibinfo
  {volume} {224}},\ \bibinfo {pages} {405} (\bibinfo {year}
  {2018})}\BibitemShut {NoStop}%
\bibitem [{\citenamefont {Xiao}\ \emph {et~al.}(2010)\citenamefont {Xiao},
  \citenamefont {Chang},\ and\ \citenamefont {Niu}}]{m2010Xiao}%
  \BibitemOpen
  \bibfield  {author} {\bibinfo {author} {\bibfnamefont {D.}~\bibnamefont
  {Xiao}}, \bibinfo {author} {\bibfnamefont {M.-C.}\ \bibnamefont {Chang}},\
  and\ \bibinfo {author} {\bibfnamefont {Q.}~\bibnamefont {Niu}},\ }\bibfield
  {title} {\bibinfo {title} {Berry phase effects on electronic properties},\
  }\href {https://doi.org/10.1103/RevModPhys.82.1959} {\bibfield  {journal}
  {\bibinfo  {journal} {Rev. Mod. Phys.}\ }\textbf {\bibinfo {volume} {82}},\
  \bibinfo {pages} {1959} (\bibinfo {year} {2010})}\BibitemShut {NoStop}%
\bibitem [{\citenamefont {Thouless}\ \emph {et~al.}(1982)\citenamefont
  {Thouless}, \citenamefont {Kohmoto}, \citenamefont {Nightingale},\ and\
  \citenamefont {den Nijs}}]{n1982Thouless}%
  \BibitemOpen
  \bibfield  {author} {\bibinfo {author} {\bibfnamefont {D.~J.}\ \bibnamefont
  {Thouless}}, \bibinfo {author} {\bibfnamefont {M.}~\bibnamefont {Kohmoto}},
  \bibinfo {author} {\bibfnamefont {M.~P.}\ \bibnamefont {Nightingale}},\ and\
  \bibinfo {author} {\bibfnamefont {M.}~\bibnamefont {den Nijs}},\ }\bibfield
  {title} {\bibinfo {title} {Quantized hall conductance in a two-dimensional
  periodic potential},\ }\href {https://doi.org/10.1103/PhysRevLett.49.405}
  {\bibfield  {journal} {\bibinfo  {journal} {Phys. Rev. Lett.}\ }\textbf
  {\bibinfo {volume} {49}},\ \bibinfo {pages} {405} (\bibinfo {year}
  {1982})}\BibitemShut {NoStop}%
\bibitem [{\citenamefont {Shen}\ \emph {et~al.}(2014)\citenamefont {Shen},
  \citenamefont {Wang}, \citenamefont {Jin}, \citenamefont {Huang},
  \citenamefont {Ying}, \citenamefont {Li}, \citenamefont {Lai}, \citenamefont
  {Zhou}, \citenamefont {Zhang}, \citenamefont {Lin}, \citenamefont {Wu},\ and\
  \citenamefont {Chen}}]{CCAShen2014}%
  \BibitemOpen
  \bibfield  {author} {\bibinfo {author} {\bibfnamefont {S.}~\bibnamefont
  {Shen}}, \bibinfo {author} {\bibfnamefont {G.}~\bibnamefont {Wang}}, \bibinfo
  {author} {\bibfnamefont {S.}~\bibnamefont {Jin}}, \bibinfo {author}
  {\bibfnamefont {Q.}~\bibnamefont {Huang}}, \bibinfo {author} {\bibfnamefont
  {T.}~\bibnamefont {Ying}}, \bibinfo {author} {\bibfnamefont {D.}~\bibnamefont
  {Li}}, \bibinfo {author} {\bibfnamefont {X.}~\bibnamefont {Lai}}, \bibinfo
  {author} {\bibfnamefont {T.}~\bibnamefont {Zhou}}, \bibinfo {author}
  {\bibfnamefont {H.}~\bibnamefont {Zhang}}, \bibinfo {author} {\bibfnamefont
  {Z.}~\bibnamefont {Lin}}, \bibinfo {author} {\bibfnamefont {X.}~\bibnamefont
  {Wu}},\ and\ \bibinfo {author} {\bibfnamefont {X.}~\bibnamefont {Chen}},\
  }\bibfield  {title} {\bibinfo {title} {Tunable cobalt vacancies and related
  properties in lacoxas2},\ }\href {https://doi.org/10.1021/cm5029639}
  {\bibfield  {journal} {\bibinfo  {journal} {Chemistry of Materials}\ }\textbf
  {\bibinfo {volume} {26}},\ \bibinfo {pages} {6221} (\bibinfo {year}
  {2014})}\BibitemShut {NoStop}%
\bibitem [{\citenamefont {Mangez}\ \emph {et~al.}(1976)\citenamefont {Mangez},
  \citenamefont {Issi},\ and\ \citenamefont {Heremans}}]{Mangez1967Bi}%
  \BibitemOpen
  \bibfield  {author} {\bibinfo {author} {\bibfnamefont {J.~H.}\ \bibnamefont
  {Mangez}}, \bibinfo {author} {\bibfnamefont {J.~P.}\ \bibnamefont {Issi}},\
  and\ \bibinfo {author} {\bibfnamefont {J.}~\bibnamefont {Heremans}},\
  }\bibfield  {title} {\bibinfo {title} {Transport properties of bismuth in
  quantizing magnetic fields},\ }\href
  {https://doi.org/10.1103/PhysRevB.14.4381} {\bibfield  {journal} {\bibinfo
  {journal} {Phys. Rev. B}\ }\textbf {\bibinfo {volume} {14}},\ \bibinfo
  {pages} {4381} (\bibinfo {year} {1976})}\BibitemShut {NoStop}%
\bibitem [{\citenamefont {Liu}\ \emph {et~al.}(2018)\citenamefont {Liu},
  \citenamefont {Sun}, \citenamefont {Kumar}, \citenamefont {Muechler},
  \citenamefont {Sun}, \citenamefont {Jiao}, \citenamefont {Yang},
  \citenamefont {Liu}, \citenamefont {Liang}, \citenamefont {Xu}, \citenamefont
  {Kroder}, \citenamefont {S{\"u}{\ss}}, \citenamefont {Borrmann},
  \citenamefont {Shekhar}, \citenamefont {Wang}, \citenamefont {Xi},
  \citenamefont {Wang}, \citenamefont {Schnelle}, \citenamefont {Wirth},
  \citenamefont {Chen}, \citenamefont {Goennenwein},\ and\ \citenamefont
  {Felser}}]{Liu2018Co3Sn2S2}%
  \BibitemOpen
  \bibfield  {author} {\bibinfo {author} {\bibfnamefont {E.}~\bibnamefont
  {Liu}}, \bibinfo {author} {\bibfnamefont {Y.}~\bibnamefont {Sun}}, \bibinfo
  {author} {\bibfnamefont {N.}~\bibnamefont {Kumar}}, \bibinfo {author}
  {\bibfnamefont {L.}~\bibnamefont {Muechler}}, \bibinfo {author}
  {\bibfnamefont {A.}~\bibnamefont {Sun}}, \bibinfo {author} {\bibfnamefont
  {L.}~\bibnamefont {Jiao}}, \bibinfo {author} {\bibfnamefont {S.-Y.}\
  \bibnamefont {Yang}}, \bibinfo {author} {\bibfnamefont {D.}~\bibnamefont
  {Liu}}, \bibinfo {author} {\bibfnamefont {A.}~\bibnamefont {Liang}}, \bibinfo
  {author} {\bibfnamefont {Q.}~\bibnamefont {Xu}}, \bibinfo {author}
  {\bibfnamefont {J.}~\bibnamefont {Kroder}}, \bibinfo {author} {\bibfnamefont
  {V.}~\bibnamefont {S{\"u}{\ss}}}, \bibinfo {author} {\bibfnamefont
  {H.}~\bibnamefont {Borrmann}}, \bibinfo {author} {\bibfnamefont
  {C.}~\bibnamefont {Shekhar}}, \bibinfo {author} {\bibfnamefont
  {Z.}~\bibnamefont {Wang}}, \bibinfo {author} {\bibfnamefont {C.}~\bibnamefont
  {Xi}}, \bibinfo {author} {\bibfnamefont {W.}~\bibnamefont {Wang}}, \bibinfo
  {author} {\bibfnamefont {W.}~\bibnamefont {Schnelle}}, \bibinfo {author}
  {\bibfnamefont {S.}~\bibnamefont {Wirth}}, \bibinfo {author} {\bibfnamefont
  {Y.}~\bibnamefont {Chen}}, \bibinfo {author} {\bibfnamefont {S.~T.~B.}\
  \bibnamefont {Goennenwein}},\ and\ \bibinfo {author} {\bibfnamefont
  {C.}~\bibnamefont {Felser}},\ }\bibfield  {title} {\bibinfo {title} {Giant
  anomalous hall effect in a ferromagnetic kagome-lattice semimetal},\ }\href
  {https://doi.org/10.1038/s41567-018-0234-5} {\bibfield  {journal} {\bibinfo
  {journal} {Nat. Phys.}\ }\textbf {\bibinfo {volume} {14}},\ \bibinfo {pages}
  {1125} (\bibinfo {year} {2018})}\BibitemShut {NoStop}%
\bibitem [{\citenamefont {Bredl}\ \emph {et~al.}(1984)\citenamefont {Bredl},
  \citenamefont {Horn}, \citenamefont {Steglich}, \citenamefont {L\"uthi},\
  and\ \citenamefont {Martin}}]{Bredl1984CeCu2Si2}%
  \BibitemOpen
  \bibfield  {author} {\bibinfo {author} {\bibfnamefont {C.~D.}\ \bibnamefont
  {Bredl}}, \bibinfo {author} {\bibfnamefont {S.}~\bibnamefont {Horn}},
  \bibinfo {author} {\bibfnamefont {F.}~\bibnamefont {Steglich}}, \bibinfo
  {author} {\bibfnamefont {B.}~\bibnamefont {L\"uthi}},\ and\ \bibinfo {author}
  {\bibfnamefont {R.~M.}\ \bibnamefont {Martin}},\ }\bibfield  {title}
  {\bibinfo {title} {Low-temperature specific heat of
  {C}e{C}${\mathrm{u}}_{2}${S}${\mathrm{i}}_{2}$ and {C}e{A}${\mathrm{l}}_{3}$:
  Coherence effects in {K}ondo lattice systems},\ }\href
  {https://doi.org/10.1103/PhysRevLett.52.1982} {\bibfield  {journal} {\bibinfo
   {journal} {Phys. Rev. Lett.}\ }\textbf {\bibinfo {volume} {52}},\ \bibinfo
  {pages} {1982} (\bibinfo {year} {1984})}\BibitemShut {NoStop}%
\bibitem [{\citenamefont {Jaccard}\ and\ \citenamefont
  {Flouquet}(1985)}]{Jaccard1985HF}%
  \BibitemOpen
  \bibfield  {author} {\bibinfo {author} {\bibfnamefont {D.}~\bibnamefont
  {Jaccard}}\ and\ \bibinfo {author} {\bibfnamefont {J.}~\bibnamefont
  {Flouquet}},\ }\bibfield  {title} {\bibinfo {title} {The normal phase of
  heavy fermion compounds},\ }\href
  {https://doi.org/https://doi.org/10.1016/0304-8853(85)90354-3} {\bibfield
  {journal} {\bibinfo  {journal} {J. Magn. Magn. Mater.}\ }\textbf {\bibinfo
  {volume} {47-48}},\ \bibinfo {pages} {45} (\bibinfo {year}
  {1985})}\BibitemShut {NoStop}%
\bibitem [{\citenamefont {Sparn}\ \emph {et~al.}(1985)\citenamefont {Sparn},
  \citenamefont {Lieke}, \citenamefont {Gottwick}, \citenamefont {Steglich},\
  and\ \citenamefont {Grewe}}]{Sparn1985Kondo}%
  \BibitemOpen
  \bibfield  {author} {\bibinfo {author} {\bibfnamefont {G.}~\bibnamefont
  {Sparn}}, \bibinfo {author} {\bibfnamefont {W.}~\bibnamefont {Lieke}},
  \bibinfo {author} {\bibfnamefont {U.}~\bibnamefont {Gottwick}}, \bibinfo
  {author} {\bibfnamefont {F.}~\bibnamefont {Steglich}},\ and\ \bibinfo
  {author} {\bibfnamefont {N.}~\bibnamefont {Grewe}},\ }\bibfield  {title}
  {\bibinfo {title} {Low-temperature transport properties of {K}ondo
  lattices},\ }\href
  {https://doi.org/https://doi.org/10.1016/0304-8853(85)90482-2} {\bibfield
  {journal} {\bibinfo  {journal} {J. Magn. Magn. Mater.}\ }\textbf {\bibinfo
  {volume} {47-48}},\ \bibinfo {pages} {521} (\bibinfo {year}
  {1985})}\BibitemShut {NoStop}%
\bibitem [{\citenamefont {Ayache}\ \emph {et~al.}(1987)\citenamefont {Ayache},
  \citenamefont {Beille}, \citenamefont {Bonjour}, \citenamefont {Calemczuk},
  \citenamefont {Creuzet}, \citenamefont {Gignoux}, \citenamefont {Najib},
  \citenamefont {Schmitt}, \citenamefont {Voiron},\ and\ \citenamefont
  {Zerguine}}]{Ayache1987CePt2Si2}%
  \BibitemOpen
  \bibfield  {author} {\bibinfo {author} {\bibfnamefont {C.}~\bibnamefont
  {Ayache}}, \bibinfo {author} {\bibfnamefont {J.}~\bibnamefont {Beille}},
  \bibinfo {author} {\bibfnamefont {E.}~\bibnamefont {Bonjour}}, \bibinfo
  {author} {\bibfnamefont {R.}~\bibnamefont {Calemczuk}}, \bibinfo {author}
  {\bibfnamefont {G.}~\bibnamefont {Creuzet}}, \bibinfo {author} {\bibfnamefont
  {D.}~\bibnamefont {Gignoux}}, \bibinfo {author} {\bibfnamefont
  {A.}~\bibnamefont {Najib}}, \bibinfo {author} {\bibfnamefont
  {D.}~\bibnamefont {Schmitt}}, \bibinfo {author} {\bibfnamefont
  {J.}~\bibnamefont {Voiron}},\ and\ \bibinfo {author} {\bibfnamefont
  {M.}~\bibnamefont {Zerguine}},\ }\bibfield  {title} {\bibinfo {title}
  {Specific heat and pressure effects in the {K}ondo lattice compound
  {CePt$_2$Si$_2$}},\ }\href
  {https://doi.org/https://doi.org/10.1016/0304-8853(87)90601-9} {\bibfield
  {journal} {\bibinfo  {journal} {J. Magn. Magn. Mater.}\ }\textbf {\bibinfo
  {volume} {63-64}},\ \bibinfo {pages} {329} (\bibinfo {year}
  {1987})}\BibitemShut {NoStop}%
\bibitem [{\citenamefont {Sato}\ \emph {et~al.}(1987)\citenamefont {Sato},
  \citenamefont {Zhao}, \citenamefont {Pratt}, \citenamefont
  {\ifmmode~\bar{O}\else \={O}\fi{}nuki},\ and\ \citenamefont
  {Komatsubara}}]{Sato1987CeCu6}%
  \BibitemOpen
  \bibfield  {author} {\bibinfo {author} {\bibfnamefont {H.}~\bibnamefont
  {Sato}}, \bibinfo {author} {\bibfnamefont {J.}~\bibnamefont {Zhao}}, \bibinfo
  {author} {\bibfnamefont {W.~P.}\ \bibnamefont {Pratt}}, \bibinfo {author}
  {\bibfnamefont {Y.}~\bibnamefont {\ifmmode~\bar{O}\else \={O}\fi{}nuki}},\
  and\ \bibinfo {author} {\bibfnamefont {T.}~\bibnamefont {Komatsubara}},\
  }\bibfield  {title} {\bibinfo {title} {Transport properties of the
  heavy-fermion compound {CeCu$_6$} down to 14 mk},\ }\href
  {https://doi.org/10.1103/PhysRevB.36.8841} {\bibfield  {journal} {\bibinfo
  {journal} {Phys. Rev. B}\ }\textbf {\bibinfo {volume} {36}},\ \bibinfo
  {pages} {8841} (\bibinfo {year} {1987})}\BibitemShut {NoStop}%
\bibitem [{\citenamefont {Amato}\ \emph {et~al.}(1988)\citenamefont {Amato},
  \citenamefont {Jaccard}, \citenamefont {Sierro}, \citenamefont {Lapierre},
  \citenamefont {Haen}, \citenamefont {Lejay},\ and\ \citenamefont
  {Flouquet}}]{Amato1988CeRu2Si2}%
  \BibitemOpen
  \bibfield  {author} {\bibinfo {author} {\bibfnamefont {A.}~\bibnamefont
  {Amato}}, \bibinfo {author} {\bibfnamefont {D.}~\bibnamefont {Jaccard}},
  \bibinfo {author} {\bibfnamefont {J.}~\bibnamefont {Sierro}}, \bibinfo
  {author} {\bibfnamefont {F.}~\bibnamefont {Lapierre}}, \bibinfo {author}
  {\bibfnamefont {P.}~\bibnamefont {Haen}}, \bibinfo {author} {\bibfnamefont
  {P.}~\bibnamefont {Lejay}},\ and\ \bibinfo {author} {\bibfnamefont
  {J.}~\bibnamefont {Flouquet}},\ }\bibfield  {title} {\bibinfo {title}
  {Thermopower and magneto-thermopower of {CeRu$_2$Si$_2$} single crystals},\
  }\href {https://doi.org/https://doi.org/10.1016/0304-8853(88)90389-7}
  {\bibfield  {journal} {\bibinfo  {journal} {J. Magn. Magn. Mater.}\ }\textbf
  {\bibinfo {volume} {76-77}},\ \bibinfo {pages} {263} (\bibinfo {year}
  {1988})}\BibitemShut {NoStop}%
\bibitem [{\citenamefont {Bhattacharjee}\ \emph {et~al.}(1989)\citenamefont
  {Bhattacharjee}, \citenamefont {Coqblin}, \citenamefont {Raki}, \citenamefont
  {Forro}, \citenamefont {Ayache},\ and\ \citenamefont
  {Schmitt}}]{Bhattacharjee1989CePt2Si2}%
  \BibitemOpen
  \bibfield  {author} {\bibinfo {author} {\bibfnamefont {A.~K.}\ \bibnamefont
  {Bhattacharjee}}, \bibinfo {author} {\bibfnamefont {B.}~\bibnamefont
  {Coqblin}}, \bibinfo {author} {\bibfnamefont {M.}~\bibnamefont {Raki}},
  \bibinfo {author} {\bibfnamefont {L.}~\bibnamefont {Forro}}, \bibinfo
  {author} {\bibfnamefont {C.}~\bibnamefont {Ayache}},\ and\ \bibinfo {author}
  {\bibfnamefont {D.}~\bibnamefont {Schmitt}},\ }\bibfield  {title} {\bibinfo
  {title} {Anisotropy of transport properties in the {K}ondo compound
  {CePt$_2$Si$_2$}: experiments and theory},\ }\href
  {https://doi.org/10.1051/jphys:0198900500180278100} {\bibfield  {journal}
  {\bibinfo  {journal} {J. Phys. France}\ }\textbf {\bibinfo {volume} {50}},\
  \bibinfo {pages} {2781} (\bibinfo {year} {1989})}\BibitemShut {NoStop}%
\bibitem [{\citenamefont {Lacerda}\ \emph {et~al.}(1989)\citenamefont
  {Lacerda}, \citenamefont {de~Visser}, \citenamefont {Haen}, \citenamefont
  {Lejay},\ and\ \citenamefont {Flouquet}}]{Lacerda1989CeRu2Si2}%
  \BibitemOpen
  \bibfield  {author} {\bibinfo {author} {\bibfnamefont {A.}~\bibnamefont
  {Lacerda}}, \bibinfo {author} {\bibfnamefont {A.}~\bibnamefont {de~Visser}},
  \bibinfo {author} {\bibfnamefont {P.}~\bibnamefont {Haen}}, \bibinfo {author}
  {\bibfnamefont {P.}~\bibnamefont {Lejay}},\ and\ \bibinfo {author}
  {\bibfnamefont {J.}~\bibnamefont {Flouquet}},\ }\bibfield  {title} {\bibinfo
  {title} {Thermal properties of heavy-fermion {CeRu$_2$Si$_2$}},\ }\href
  {https://doi.org/10.1103/PhysRevB.40.8759} {\bibfield  {journal} {\bibinfo
  {journal} {Phys. Rev. B}\ }\textbf {\bibinfo {volume} {40}},\ \bibinfo
  {pages} {8759} (\bibinfo {year} {1989})}\BibitemShut {NoStop}%
\bibitem [{\citenamefont {L\"ohneysen}\ \emph {et~al.}(1994)\citenamefont
  {L\"ohneysen}, \citenamefont {Pietrus}, \citenamefont {Portisch},
  \citenamefont {Schlager}, \citenamefont {Schr\"oder}, \citenamefont {Sieck},\
  and\ \citenamefont {Trappmann}}]{Lohneysen1994HF}%
  \BibitemOpen
  \bibfield  {author} {\bibinfo {author} {\bibfnamefont {H.~v.}\ \bibnamefont
  {L\"ohneysen}}, \bibinfo {author} {\bibfnamefont {T.}~\bibnamefont
  {Pietrus}}, \bibinfo {author} {\bibfnamefont {G.}~\bibnamefont {Portisch}},
  \bibinfo {author} {\bibfnamefont {H.~G.}\ \bibnamefont {Schlager}}, \bibinfo
  {author} {\bibfnamefont {A.}~\bibnamefont {Schr\"oder}}, \bibinfo {author}
  {\bibfnamefont {M.}~\bibnamefont {Sieck}},\ and\ \bibinfo {author}
  {\bibfnamefont {T.}~\bibnamefont {Trappmann}},\ }\bibfield  {title} {\bibinfo
  {title} {Non-fermi-liquid behavior in a heavy-fermion alloy at a magnetic
  instability},\ }\href {https://doi.org/10.1103/PhysRevLett.72.3262}
  {\bibfield  {journal} {\bibinfo  {journal} {Phys. Rev. Lett.}\ }\textbf
  {\bibinfo {volume} {72}},\ \bibinfo {pages} {3262} (\bibinfo {year}
  {1994})}\BibitemShut {NoStop}%
\bibitem [{\citenamefont {Bianchi}\ \emph {et~al.}(2003)\citenamefont
  {Bianchi}, \citenamefont {Movshovich}, \citenamefont {Vekhter}, \citenamefont
  {Pagliuso},\ and\ \citenamefont {Sarrao}}]{Bianchi2003CeCoIn5}%
  \BibitemOpen
  \bibfield  {author} {\bibinfo {author} {\bibfnamefont {A.}~\bibnamefont
  {Bianchi}}, \bibinfo {author} {\bibfnamefont {R.}~\bibnamefont {Movshovich}},
  \bibinfo {author} {\bibfnamefont {I.}~\bibnamefont {Vekhter}}, \bibinfo
  {author} {\bibfnamefont {P.~G.}\ \bibnamefont {Pagliuso}},\ and\ \bibinfo
  {author} {\bibfnamefont {J.~L.}\ \bibnamefont {Sarrao}},\ }\bibfield  {title}
  {\bibinfo {title} {Avoided antiferromagnetic order and quantum critical point
  in {CeCoIn$_5$}},\ }\href {https://doi.org/10.1103/PhysRevLett.91.257001}
  {\bibfield  {journal} {\bibinfo  {journal} {Phys. Rev. Lett.}\ }\textbf
  {\bibinfo {volume} {91}},\ \bibinfo {pages} {257001} (\bibinfo {year}
  {2003})}\BibitemShut {NoStop}%
\bibitem [{\citenamefont {Bel}\ \emph {et~al.}(2004)\citenamefont {Bel},
  \citenamefont {Behnia}, \citenamefont {Nakajima}, \citenamefont {Izawa},
  \citenamefont {Matsuda}, \citenamefont {Shishido}, \citenamefont {Settai},\
  and\ \citenamefont {Onuki}}]{Bel2004CeCoIn5}%
  \BibitemOpen
  \bibfield  {author} {\bibinfo {author} {\bibfnamefont {R.}~\bibnamefont
  {Bel}}, \bibinfo {author} {\bibfnamefont {K.}~\bibnamefont {Behnia}},
  \bibinfo {author} {\bibfnamefont {Y.}~\bibnamefont {Nakajima}}, \bibinfo
  {author} {\bibfnamefont {K.}~\bibnamefont {Izawa}}, \bibinfo {author}
  {\bibfnamefont {Y.}~\bibnamefont {Matsuda}}, \bibinfo {author} {\bibfnamefont
  {H.}~\bibnamefont {Shishido}}, \bibinfo {author} {\bibfnamefont
  {R.}~\bibnamefont {Settai}},\ and\ \bibinfo {author} {\bibfnamefont
  {Y.}~\bibnamefont {Onuki}},\ }\bibfield  {title} {\bibinfo {title} {Giant
  nernst effect in {CeCoIn$_5$}},\ }\href
  {https://doi.org/10.1103/PhysRevLett.92.217002} {\bibfield  {journal}
  {\bibinfo  {journal} {Phys. Rev. Lett.}\ }\textbf {\bibinfo {volume} {92}},\
  \bibinfo {pages} {217002} (\bibinfo {year} {2004})}\BibitemShut {NoStop}%
\end{thebibliography}%

\end{document}